\PassOptionsToPackage{dvipsnames,table,xcdraw}{xcolor}
\documentclass[acmlarge,nonacm]{acmart}
\usepackage{tikz}
\PassOptionsToPackage{dvipsnames,table,xcdraw}{xcolor}
\usepackage[normalem]{ulem}
\usepackage{array}
\usepackage{wrapfig}
\newcommand{\blue}[1]{\textcolor{black}{#1}}
\newcommand{\revise}[1]{\textcolor{black}{#1}}
\AtBeginDocument{%
  }

\acmJournal{IMWUT} % Journal name (abbreviated title will be formatted correctly)
\acmVolume{0}      % Placeholder for volume
\acmNumber{0}      % Placeholder for issue
\acmArticle{0}     % Placeholder for article number
\acmYear{2024}     % Update as needed
\acmMonth{1}       % Update as needed

\acmDOI{10.1145/0000000.0000000} % Use correct DOI or placeholder

\acmISBN{978-1-4503-XXXX-X/2018/06}
\usepackage[most]{tcolorbox}
\usepackage{url} % for \url if not already included
\usepackage{float}
\usepackage{caption}
\usepackage{subcaption}
\usepackage{graphicx}
\usepackage{array}
\usepackage{booktabs}  
\usepackage{tabularx} 
\usepackage{pdflscape}
\usepackage[table]{xcolor}
\usepackage{xcolor}
\usepackage[normalem]{ulem} % for \ULon
\usepackage{longtable}
\usepackage{multirow}
\usepackage{makecell}
\usepackage[utf8]{inputenc}
\usepackage{hyperref}
\usepackage{fancyhdr}
\usepackage{url}
\usepackage[dvipsnames]{xcolor}
\renewcommand{\arraystretch}{1.3}
\newcolumntype{L}[1]{>{\raggedright\arraybackslash}p{#1}}
\newcolumntype{C}[1]{>{\centering\arraybackslash}p{#1}}

\newcommand{\system}{GIDEA}
\tcbset{quote style/.style={
  enhanced,
  colback=gray!5,
  colframe=black,
  fonttitle=\bfseries,
  boxrule=0.4pt,
  arc=0pt,
  left=6pt,
  right=6pt,
  top=4pt,
  bottom=4pt,
  before skip=5pt,
  after skip=5pt
}}

\begin{document}

\title{Design and Evaluation of Generative Agent-based Platform for Human-Assistant Interaction Research: A Tale of 10 User Studies}

\author{Ziyi Xuan}
\affiliation{\institution{Lehigh University} \country{USA}}
\email{zix222@lehigh.edu}

\author{Yiwen Wu}
\affiliation{\institution{Lehigh University} \country{USA}}
\email{yiw423@lehigh.edu}

\author{Xuhai Xu}
\affiliation{\institution{Columbia University} \country{USA}}
\email{xx2489@cumc.columbia.edu}

\author{Vinod Namboodiri}
\affiliation{\institution{Lehigh University} \country{USA}}
\email{vin423@lehigh.edu}

\author{Mooi Choo Chuah}
\affiliation{\institution{Lehigh University} \country{USA}}
\email{mcc7@lehigh.edu}

\author{Yu Yang}
\affiliation{\institution{Lehigh University} \country{USA}}
\email{yuyang@lehigh.edu}

\begin{abstract}
Designing and evaluating personalized and proactive assistant agents remains challenging due to the time, cost, and ethical concerns associated with human-in-the-loop experimentation. Existing Human-Computer Interaction (HCI) methods often require extensive physical setup and human participation, which introduces privacy concerns and limits scalability. Simulated environments offer a partial solution but are typically constrained by rule-based scenarios and still depend heavily on human input to guide interactions and interpret results.
Recent advances in large language models (LLMs) have introduced the possibility of generative agents that can simulate realistic human behavior, reasoning, and social dynamics. 
However, their effectiveness in modeling human-assistant interactions remains largely unexplored.
To address this gap, we present a generative agent-based simulation platform designed to simulate human-assistant interactions. We identify ten prior studies on assistant agents that span different aspects of interaction design and replicate these studies using our simulation platform.
Our results show that fully simulated experiments using generative agents can approximate key aspects of human-assistant interactions. Based on these simulations, we are able to replicate the core conclusions of the original studies. Our work provides a scalable and cost-effective approach for studying assistant agent design without requiring live human subjects.
\revise{Additional resources and project materials are available at \url{https://dash-gidea.github.io/}.}

\end{abstract}


%  \includegraphics[width=\textwidth]{figure/teaser.png}
%  \caption{}
%  \Description{}

\received{20 February 2007}
\received[revised]{12 March 2009}
\received[accepted]{5 June 2009}

\maketitle

\section{Introduction}

Designing intelligent assistants that can proactively support human needs has become a long-standing goal in human–computer interaction (HCI) research~\cite{luna2024kdd}. 
With user consent, such systems are expected to infer user states, predict needs, and deliver timely, context-aware support. 
To better design and improve these systems, researchers have studied human-assistant interactions through controlled experiments, using customer-built prototypes or through platforms such as Amazon Alexa and Google Assistant~\cite{amazon_alexa, google_assistant, bentley2018understanding, minder2023voice}. 
However, conducting experiments in physical environments is often labor-intensive, time-consuming, and logistically complex. Practical challenges—such as setting up experimental environments, managing participants, and collecting multi-source interaction data—limit the scale, flexibility, and personalization of study designs, particularly for longitudinal deployments. Moreover, privacy and ethical considerations necessitate additional safeguards to protect participants, which in turn limit the types of interactions and scenarios that can be explored in live deployments.

% To address these challenges, researchers have explored simulator-based approaches to simulate a smart environment and activities of daily living (ADL). 
To address these barriers, researchers have explored simulation-based approaches that replicate user behaviors and environmental dynamics in virtual spaces.
Platforms such as OpenSHS~\cite{alshammari2017openshs} and VirtualHome~\cite{puig2018virtualhome} offer structured frameworks to model smart environments and Activities of Daily Living (ADL), which enable algorithm testing and scenario modeling in controlled, repeatable settings~\cite{synnott2015simulation}. 
However, most existing simulation platforms are built on either manual operation or task-oriented models, which require extensive scripting and lack the flexibility to capture the variability of human behavior and physical environments. As a result, they fall short for studying human–assistant interactions, which demand personalized, adaptive, and socially complex behaviors.

Recent advances in generative AI, particularly large language models (LLMs), offer a new opportunity to enable more resource-efficient simulation approaches.
LLM-based generative agents—computational entities designed to emulate believable human behaviors~\cite{park2023generative}—can generate contextually rich language, reason over multiple turns of interaction, adapt to persona attributes, and express diverse communication styles~\cite{park2023generative, wang2024survey}. 
Building on these capabilities, we explore how generative agents can be leveraged to simulate human–assistant interactions. Specifically, we aim to understand how generative agent-simulated humans can contribute to human–assistant interaction research, and how the generated results compare to findings from existing human subject studies. 
Realizing this vision introduces several practical challenges, including data limitations, variation in research setups, and restricted access to detailed behavioral records.

\noindent \textbf{Challenge 1}: Public datasets that link human activities to internal cognitive or contextual states are limited. Existing data sources~\cite{cook_casas_2013, alshammari2017openshs, arrotta2021marble} are often small in scale and lack semantic annotations or user-centric labels, which makes it difficult to be used to evaluate the realism of simulated behaviors directly. To overcome this limitation, we identified ten well-documented human-assistant interaction studies and aligned our simulations with their experimental protocols. These studies serve as reference points for evaluating simulation quality via behavioral metrics (e.g., strategy usage, acceptance rates) and the semantic similarity of generated responses to original research findings. 

\noindent \textbf{Challenge 2}: Designing a platform that can accommodate the wide variation in research setups across human–assistant interaction studies (e.g., target participants, interaction techniques) is challenging. A unified format and method are needed to translate diverse study designs into consistent, operable guidelines that replace the traditional guiding role of human researchers and support automated simulation. To meet this need, we developed a modular system design that accommodates variations across all aspects of experimental protocols. The information necessary for autonomously guiding a study—such as questionnaires, task sequences, interaction strategies, and experimental rules—is encoded into structured, reusable templates. We also designed a simulation workflow that dynamically prompts the generative agent with this configured information, including an interaction memory, allowing it to manage each phase of the study and generate context-appropriate responses throughout the simulation process. 

\noindent \textbf{Challenge 3}: Many prior HCI studies do not release full interaction logs due to privacy concerns, limiting the availability of fine-grained behavioral data for replication. To address this challenge, we carefully design an evaluation methodology that aligns simulated behaviors with reported qualitative descriptions and interaction patterns from the original studies to enable meaningful comparison in the absence of detailed logs.

% While this restricts the completeness of available evidence, we mitigate the gap by carefully designing our evaluation based on qualitative descriptions and reported behavioral patterns in the original papers to guide simulation setup and evaluation.

%Third, prior human–assistant interaction studies generally do not release any recorded interaction data due to privacy concerns. As a result, the information available in original user studies is often limited or selectively reported, which restricts the depth and completeness of the evidence we can extract. This poses challenges for accurately reconstructing participant behaviors or responses during simulation.

%More importantly, we present the first comprehensive evaluation of generative agent–based simulation for human–assistant interactions. We replicate ten prior user studies spanning various themes such as proactive assistance, interruptibility, adaptive personalization, and user engagement. We demonstrate that generative agents in \system{} can reproduce key behavioral patterns observed in human-subject studies and provide a flexible way to explore additional study conditions, reducing the burden in early-stage design before involving actual human participants. 

In this work, we present \system{}, the first comprehensive evaluation of generative agent–based simulation for human–assistant interactions. Grounded on the existing research workflow (details in Fig.\ref{fig:comparison_trad_LLM_HCI}), %\system{} provides a modular, extensible framework that enables researchers to define experimental conditions and automatically instantiate virtual avatars. 
we replicate ten prior user studies spanning various themes such as proactive assistance, interruptibility, adaptive personalization, and user engagement. We demonstrate that generative agents in \system{} can reproduce key behavioral patterns observed in human-subject studies and provide a flexible way to explore additional study conditions, which potentially reduces the burden in early-stage design before involving actual human participants. 
%These avatars interact with an assistant agent in real time, adapting their behaviors based on memory, context, and dialogue history. Compared to the existing simulation platforms (Table~\ref{tab:gidea_comparison}), \system{} provides the following advantages that extend beyond prior work: %(change the following to the advantages)
%Compared to existing human activity simulation platforms designed to support HCI and system research, \system{} provides a structured and extensible environment that replicates human–assistant interactions with enhanced experimental control, memory-driven behavior modeling, and autonomous execution. The following advantages extend beyond the capabilities of prior platforms:
We summarize our contributions as follows:

\begin{itemize} 
\item We present \system{}, a general-purpose simulation platform for replicating human-assistant interaction experiments using LLM-based generative agents to reenact human behaviors. The platform models participants, environments, and interactions, enabling reproducible simulation of complex user studies.

\item We introduce a unified method to translate diverse experimental protocols into LLM-compatible prompts. This prompting strategy supports a wide range of experimental setups and tasks, allowing \system{} to be applied across various domains within human-assistant interaction research, including smart homes, voice assistants, and decision-support systems.

\item We replicate ten published human-assistant interaction experiments using \system{}, where we observe that the generative agent produces meaningful, human-like responses that offer valuable insights for researchers. These responses closely align with original human-subject results, validating the fidelity and generalizability of generative agent-based behavioral simulation in controlled study settings. We will open source both the platform and collected results from the experiments.

%\item We develop a Unity3D-based embodied simulation environment that integrates with \system{}, enabling the capture of structured spatial context. This information is incorporated into generative agents’ prompting to support the generation of context-aware interactions during simulation.

%\item We validate the ability of LLM agents to simulate participants with coherent, goal-aligned attitudes that reflect the overarching intent and structure of HCI experiments. These agents generate meaningful, human-like responses that offer valuable insights for researchers—supporting experiment design, refinement, and pre-study evaluation.
\end{itemize}

\section{Related Work}

% \textcolor{red}{List limitations/challenges of current HCI, particularly IPA interactions. Propose that we can use simulators to minimize human efforts and improve efficiency.}

\subsection{Design and Evaluation of Intelligent Assistants Agent}

% \textcolor{red}{research topics, moved from previous pilot study section} 
\textbf{Design Focus:} The design of intelligent assistants for everyday use increasingly emphasizes personalized, context-aware, and adaptive interactions that align with user expectations and support ongoing activities. Effective interaction design requires a careful balance between automation and user control, shaped by how users perceive, engage with, and configure assistant behaviors.
One important theme in prior work is the importance of personalization in tone, interaction style, and decision-making strategies~\cite{zargham_i_2022, clark_what_2019, chin2024like}. Personalization not only improves usability but also helps build trust, especially as users often anthropomorphize assistants, attributing social roles or intentions that influence their interpretation of system behavior~\cite{chin2024like}.
Another key perspective focuses on proactive engagement. Studies have examined how assistants initiate interactions based on context and timing~\cite{zargham_understanding_2022, wei2021understanding, cha_hello_2020}. While users generally value proactive support, they prefer subtle or permission-seeking cues, such as soft chimes, when assistants interrupt ongoing tasks~\cite{cha_hello_2020, reicherts_may_2021}. These findings point to the importance of adaptive timing strategies that maintain user agency and minimize disruption.
Finally, recent research explores user-driven approaches to configuring assistant behaviors. In-situ programming methods allow users to define or refine actions during actual use, promoting greater transparency and flexibility~\cite{oh_better_2024, liu_understanding_2023}. These approaches support a more collaborative model of interaction, where the assistant becomes a configurable partner rather than a fixed service.

% Prior work~\cite{zargham2022want, clark_what_2019, chin2024like} has emphasized the importance of personalization in tone, style, and interaction strategies to build trust and enhance usability. Users may even anthropomorphize assistants, assigning social roles or metaphors that influence how they interpret and engage with the system~\cite{chin2024like}. Beyond personalization, researchers have examined how assistants proactively engage users based on situational context and timing~\cite{zargham_understanding_2022, wei2021understanding, cha_hello_2020}. While proactivity is valued, users generally prefer subtle or permission-seeking cues—such as soft chimes—when assistants interrupt ongoing activities~\cite{cha_hello_2020, reicherts_may_2021}. The tolerance for such proactive prompts varies across individuals, reinforcing the importance of adaptive timing strategies that support user agency and maintain a sense of control. Recent studies have begun to explore more flexible user-driven mechanisms for configuring assistant behaviors. For example, \cite{oh_better_2024, liu_understanding_2023} investigate in-situ programming techniques that allow users to define or adjust assistant actions during actual use, supporting a more transparent and collaborative model of interaction.

%study_mode_group = {
%    'Storyboard': ["CS 1", "CS 6", "CS 7"],
%    'Interview': ["CS 8", "CS 9"],
%    'WoZ': ["CS 10", "CS 7", "CS 2", "CS 4", "CS 3"]
%}

\noindent\textbf{Evaluation Method:} To evaluate these systems and interaction designs, intelligent assistant research typically draws on four methodological approaches: 
\textit{in-the-wild studies}~\cite{bentley2018understanding, adaimi2021, Garg2020}, 
\textit{customized prototypes}~\cite{liu_understanding_2023, wei2021understanding}, 
\textit{interviews}~\cite{clark_what_2019, dubiel2024hey}, \textit{storyboards}~\cite{zargham_i_2022, reicherts_may_2021, wei2021understanding}, 
and \textit{Wizard-of-Oz (WoZ) simulations}~\cite{oh_better_2024, cha_hello_2020, chin2024like}. 
In-the-wild deployments provide insights into real-world usage and long-term adaptation but are resource-intensive and difficult to scale. Customized prototypes, often built on top of commercial platforms, enable the testing of new interaction strategies but require sustained development and longitudinal testing. Interviews, surveys, and storyboards allow for efficient elicitation of user feedback in early-stage design, but they lack contextual grounding and real-time user reactions. WoZ simulations enable researchers to mimic assistant intelligence without requiring full implementation, offering a middle ground for exploring speculative features and interaction dynamics. %Most experiments incorporate some form of experience sampling (ESM) to collect feedback linked to user context. While ESM can surface meaningful insights about user perception and behavior, it remains labor-intensive and limited in scalability. %These methodological challenges motivate the use of simulation-based approaches, where large language models (LLMs) can emulate intelligent assistant behavior and user response. Simulated environments allow researchers to iterate on design strategies and evaluate proactive, personalized interactions in a controlled yet dynamic setting, reducing human effort while maintaining ecological validity.

\subsection{Simulation Platforms and Datasets in Designing Intelligent Assistants}
Simulation platforms such as SIMACT and OpenSHS~\cite{bouchard2010simact, alshammari2017openshs, liu2023training, synnott2012ie} have contributed to smart environment modeling and human activity dataset generation for intelligent assistants. Habitat~\cite{habitat19iccv, szot2021habitat, puig2023habitat} and AI2-THOR~\cite{kolve2017ai2} extend this line of work by providing scalable, photorealistic 3D environments that support embodied agents in navigation, interaction, and manipulation tasks. VirtualHome~\cite{puig2018virtualhome} complements these platforms by introducing a programmatic representation of household activities, where agents execute sequences of atomic actions derived from natural language to simulate complex daily tasks.

Despite these advances, existing platforms face several limitations. While systems such as VirtualHome and OpenSHS aim to generate labeled datasets for machine learning, they remain bound by fixed scenarios and handcrafted rules, limiting their ability to simulate emergent or evolving behaviors. Scripted simulations (e.g., SIMACT, VirtualHome) provide reproducibility but restrict behavioral variability and adaptability, which are essential for studying dynamic assistive systems. Avatar-controlled platforms (e.g., IE Sim, OpenSHS) enable more naturalistic behavior but are labor-intensive and difficult to scale. Most platforms do not incorporate cognitive processes such as goal reasoning, memory decay, or error-prone decision-making. MASSHA~\cite{kamara2017massha} introduces a belief-desire-intention (BDI) reasoning model but does not support dynamic learning from interaction. These limitations highlight the need for simulation frameworks that support scalable, autonomous agents capable of generating human like activities and responses to evolving contexts and interactions.

\subsection{Large Language Model-based Agents}
Large language model (LLM)-based agents have emerged as powerful tools for simulating human cognition and behavior across a wide range of domains, including cognitive modeling, human–computer interaction, social simulation, autonomous planning, and multi-agent collaboration~\cite{park2023generative, socialmind, sumers2023cognitive, huang2024understanding,  yan2024language, jiang2025hcride, tan2024human}. These agents demonstrate capabilities such as multi-step reasoning, goal-directed decision-making, dynamic preference adjustment, and coherent logic flow in complex environments~\cite{shinn2023reflexion, he2024afspp, yang2024talk2care, tan2025llm, 10.1145/3643540}. Prior studies show that LLMs can simulate strategic responses to evolving scenarios and exhibit behaviors aligned with human-like planning, moral reasoning, and self-reflection~\cite{chiu2024dailydilemmas, an2024golden}. Expanding on these cognitive capacities, LLM-based simulations have been used to generate conversational data for social science analysis and to explore social patterns and group behavior dynamics~\cite{bail2024can, ziems2024can, wei2022emergent}. Park et al.\cite{park2024generative}, presented a multi-agent framework that replicates human behavior in domains such as policymaking and social interaction, grounded in the Big Five personality traits~\cite{john1999big}. More works further explored the integration of psychological constructs—such as Maslow’s hierarchy of needs~\cite{mcleod2007maslow} and the Ten-Item Personality Inventory (TIPI)\cite{gosling2003very}—to personalize agent behavior~\cite{cerqueiramotif, li2024quantifying}. From a cognitive modeling perspective, recent systems have incorporated emotion, memory, and social values to enrich agent behavior, while others dynamically adapt to changing environmental and physical contexts~\cite{wang2025chatgpt}. Collectively, these advancements support the development of LLM-based agents capable of simulating interactive, adaptive, and human-aligned behavior. Building on the success of LLM-based agents, our work focuses on developing a simulation framework to enable efficient and automated research on human–assistant interactions.

\begin{figure}[h]
    \centering
    \includegraphics[width=1\linewidth]{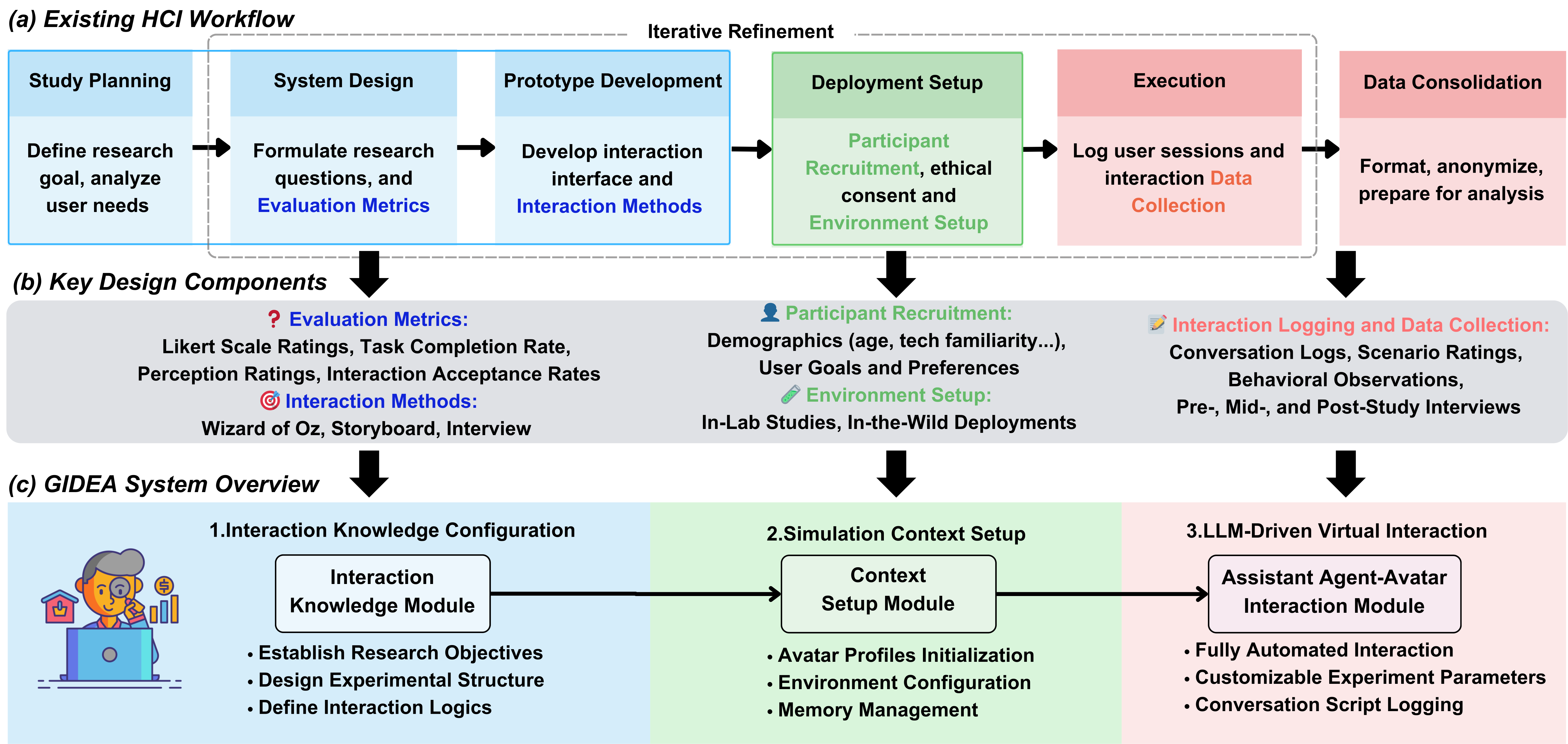}
    \vspace{-15pt}
    \caption{(a) Existing HCI studies follow a resource-intensive workflow from design to deployment and data collection. (b) Core components span evaluation metrics, assistant interfaces, participant traits, and physical environments. (c) The proposed LLM-driven framework enables efficient simulation through automated assistant agent and avatar interaction modeling.}
    \label{fig:comparison_trad_LLM_HCI}
    \vspace{-15pt}
\end{figure}

\section{Background and Motivation}
\subsection{Human–Assistant Interaction Research Workflow and Limitations}

Human–assistant interaction research typically requires long-term experimentation and extended data collection periods. Fig.~\ref{fig:comparison_trad_LLM_HCI}(a) illustrates a typical workflow, which consists of stages from study planning to data consolidation. While this established workflow has proven effective, each stage faces specific limitations that collectively motivate our simulation-based approach.

\noindent\blue{\textbf{Workflow Steps and Associated Limitations:}}

\begin{enumerate}
    \item \blue{\textbf{Study Planning:}} The process begins with study planning, where researchers define research goals based on user needs (e.g., providing timely reminders for scheduled events) and formulate specific research questions (e.g., ``Can proactive voice assistants improve user punctuality?'').

    \item \blue{\textbf{System Design:}} In the system design phase, researchers develop evaluation metrics and plan the interaction mechanisms necessary to address their research questions. Common evaluation methods include Likert-scale ratings, task completion rates, and qualitative analyses of user feedback.

    \blue{\textbf{- Limitation 1 - Inflexible Configuration Requirements:} Studies differ in user traits, interaction methods, environmental contexts, and evaluation metrics, but current approaches require extensive manual reconfiguration and system modifications to adapt experimental setups across different study conditions.}

    \item \blue{\textbf{Prototype Development:}} During prototype development, researchers operate the study through interactive systems or simulated interactions, employing methods such as prototyping, interviews, storyboards, and Wizard-of-Oz (WoZ) simulations. While prototyping enables exploration of novel designs, it can be resource-intensive, difficult to modify after deployment, and sometimes fails to elicit authentic engagement when using static or scripted approaches.

    \blue{\textbf{- Limitation 2 - Limited Scenario Reusability and Standardization:} HCI experiments often rely on standardized scenarios and interaction flows (e.g., morning preparations, proactive interruptions, task reminders), but these routines must be reimplemented for each study, leading to inconsistencies and preventing systematic comparison across research efforts.}

    \blue{\textbf{- Limitation 3 - Static and Scripted Experimental Environments:} Existing simulation platforms often rely on static or scripted scenarios that fail to capture the dynamic, temporal, and contextual changes that characterize real-world human-assistant interactions.}

    \item \blue{\textbf{Deployment Setup:}} The deployment setup phase involves obtaining ethical approval, recruiting participants, and configuring the physical environment for data collection, often requiring manual installation of sensors or smart devices. Participant recruitment is often time-consuming and subject to privacy constraints, particularly when targeting diverse demographic groups.

    \blue{\textbf{- Limitation 4 - Resource-Intensive Deployment and Execution:} Participant recruitment is time-consuming, deployment setup requires manual installation of sensors or devices, and privacy considerations limit the types of interactions that can be explored, collectively restricting the scale and flexibility of study designs.}
    
    \item \blue{\textbf{Execution:}} The execution phase carries out the study, either in controlled laboratory settings or in-the-wild environments. Each setting presents trade-offs: laboratory studies may limit ecological validity, while field deployments introduce challenges in environmental variability and long-term engagement.

    \blue{\textbf{- Limitation 5 - Lack of Psychologically Grounded Participant Modeling:} Current approaches often fail to systematically model participant personality traits and behavioral patterns, despite evidence that these factors significantly influence user responses in human-assistant interaction studies.}
    
    \item \blue{\textbf{Data Consolidation:}} Finally, the data consolidation phase formats, anonymizes, and prepares collected interaction logs for subsequent analysis. Manual transcription and coding are common but labor-intensive, potentially introducing inconsistencies that affect traceability.

    \blue{\textbf{- Limitation 6 - Poor Reproducibility and Limited Traceability:} Manual transcription and coding are labor-intensive and introduce inconsistencies. Moreover, detailed interaction logs are often unavailable due to privacy constraints, making it difficult to reproduce studies or conduct transparent post-hoc analysis.}
    
\end{enumerate}

Fig.~\ref{fig:comparison_trad_LLM_HCI}(b) further summarizes key designable components across the workflow, including evaluation metrics, interaction methods, participant recruitment criteria, environmental setup strategies, and data collection methods.
These components are critical to shaping study outcomes but also contribute to the overall complexity and resource demands of the research process.
\blue{Our platform focuses on addressing these workflow-specific limitations by leveraging generative AI to support more scalable and flexible human–assistant interaction research.}

\subsection{Opportunities with Generative AI}

We leverage the social simulation capabilities of generative AI to provide an alternative to existing human–assistant interaction experiments. Building on the conventional workflow summarized in Fig.~\ref{fig:comparison_trad_LLM_HCI}(a), our simulation framework (Fig.~\ref{fig:comparison_trad_LLM_HCI}(c)) mirrors the structure of prior research while introducing generative components to replace or augment core stages.

%experimental questions, questionnaire content, and assistant/avatar role definitions—
Specifically, large language models (LLMs) are integrated into the assistant–avatar interaction phase of the workflow. Foundational study parameters—such as research objectives, study methods, evaluation metrics, and the interaction logic between the assistant agent and avatars—are manually specified through a structured interaction Knowledge configuration (Fig.\ref{fig:comparison_trad_LLM_HCI}(c).1), providing the backbone for replicating prior studies or designing new experiments. In the simulation context setup phase (Fig.\ref{fig:comparison_trad_LLM_HCI}(c).2), researchers prepare avatar personality profiles, environmental configurations, and memory initialization, which can be manually defined or optionally generated via LLMs. Once the setup is complete, the LLM-driven virtual interaction module (Fig.~\ref{fig:comparison_trad_LLM_HCI}(c).3) autonomously simulates assistant–avatar dialogues. At this stage, generative models produce coherent, context-aware, multi-turn interactions that emulate realistic user responses and assistant behaviors, effectively virtualizing the data collection phase of human-assistant interaction research.

We develop \system{} as a modular simulation framework that integrates these components into a cohesive system. By aligning each simulation stage with its counterpart in existing experimentation, \system{} enables researchers to systematically study intelligent assistant behaviors, compare interaction strategies, and iterate on experimental designs without the resource constraints of physical deployments.
This framework provides a principled foundation for scalable, flexible, and traceable human–assistant interaction research.

\section{System Design} 

\subsection{Design Goals}
\system{} is designed to support scalable, flexible, and reproducible simulation-based human–assistant interaction research. \blue{To achieve this, we identify five key design goals derived from two sources: (1) a systematic analysis of limitations in current HCI experimentation methods as outlined in Section 3.1, and (2) technical requirements for building realistic generative agent simulations.}

\begin{itemize}
    \item \textbf{Seamless configuration of user, environment, and assistant models.}
    \blue{This goal addresses \textbf{Limitation 1} (Inflexible Configuration Requirements) and contributes to solving \textbf{Limitation 6} (Resource-Intensive Deployment) by eliminating the need for physical environment setup and participant recruitment through computational simulation. \system{} enables researchers to flexibly adapt simulations to different designs without modifying system internals.}
    
    \item \textbf{Scenario-based study design through reusable templates.}  
    \blue{This goal emerges from \textbf{Limitation 2} (Limited Scenario Reusability and Standardization), recognizing that HCI experiments often rely on standardized scenarios and interaction flows, such as morning preparations, proactive interruptions, or task reminders.} \system{} abstracts these routines into modular templates with predefined roles, triggers, and actions. Researchers can define targeted interaction scenarios and easily customize them by adjusting conditions, actions, or conversational flows without reimplementing core simulation logic.
    
    \item \textbf{Simulation of dynamic temporal and contextual changes.}  
    Real-world human-assistant interactions are triggered by evolving contexts over time. \blue{This directly addresses \textbf{Limitation 3} (Static and Scripted Experimental Environments) by moving beyond static or scripted scenarios.} \system{} models temporal progressions (e.g., time of day changes) and environmental state shifts (e.g., location, attention, device status), enabling agents to operate within dynamic, realistic environments during simulation runs.

    \item \textbf{Baseline human-aligned behavioral modeling.}  
    To promote coherent interaction patterns, \system{} assigns each simulated avatar a personality profile based on Ten-Item Personality Inventory (TIPI)~\cite{gosling2003very} scores and defined narrative traits. \blue{This goal responds to \textbf{Limitation 4} (Lack of Psychologically Grounded Participant Modeling), as personality traits significantly influence user responses in human-assistant interaction studies.} While the current implementation uses lightweight, static profiles, the modular architecture is designed to support integration with external cognitive, affective, or memory-based behavioral models in future extensions.

    \item \textbf{Traceability of simulation workflows.}  
    \system{} logs experimental parameters, environmental contexts, and assistant agent–avatar interaction histories in structured textual formats. \blue{This goal addresses \textbf{Limitation 5} (Poor Reproducibility and Limited Traceability) where detailed interaction logs are often unavailable due to privacy constraints} These detailed records support transparent post-hoc analysis, enable reconstruction of study setups, and facilitate cross-comparison across different simulation runs, even when model outputs vary due to inherent stochasticity.
\end{itemize}

% Together, these goals enable \system{} to serve as a flexible, extensible, and reproducible platform for advancing human–assistant interaction research through simulation-based experimentation.

\subsection{Module Descriptions}

\system{} operates through three coordinated modules that together support the design, configuration, and execution of simulated human--assistant interactions: 
the \textit{Interaction Knowledge Module}, which encodes the study design and logic; the \textit{Context Setup Module}, which defines avatar and environment configurations; and the \textit{Assistant Agent--Avatar Interaction Module}, which governs runtime interaction dynamics (Fig.~\ref{fig:comparison_trad_LLM_HCI}(c)). 
In the following paragraphs, we present the overall design of each module.

\noindent\blue{\textbf{Primary Model Selection:} We selected GPT-4o as the primary language model for \system{} to establish a baseline implementation. This choice was guided by its general-purpose conversational capabilities, established use in HCI research, and API stability for reproducible experiments. While our framework is model-agnostic and can accommodate future language models, GPT-4o serves as an effective proof-of-concept for validating the overall simulation pipeline.}

\subsubsection{Interaction Knowledge Module}
This module defines the structure and logic of each simulation and is initialized at the beginning of each case study (An example of interaction knowledge is shown in Appendix~\ref{appendix:interaction_knowledge}).

\noindent\textbf{Purpose:} It serves as the foundation of each simulation, capturing the experimental structure and assistant agent–avatar interaction logic that define a given simulation instance.

\noindent\textbf{Configuration:} Researchers begin by defining their study goals and then configure the module accordingly. They populate it with study-relevant metadata, such as evaluation metrics, which may be qualitative (e.g., pre-, mid-, and post-study interview questions) or quantitative (e.g., Likert-scale ratings, task completion rates)—all commonly used in human–assistant interaction studies. Depending on the research scenario, the assistant agent may act as a study facilitator or proactive assistant, while the avatar assumes the role of a simulated participant. 

\noindent\textbf{Role Separation:} To preserve realism and prevent knowledge leakage, the assistant agent and avatar are configured with asymmetric knowledge contexts. The assistant agent has access to the full case study setup, including the study instructions, configured evaluation metrics, and predefined logic for how it should behave during the simulation. In contrast, avatars are restricted to participant-facing information, such as environmental context and interaction history.

\noindent\textbf{Implementation:} The full simulation specification is encoded in a structured JSON schema, distilled from prior studies and interviews with HCI researchers. This schema supports sanity checking, reproducibility, and consistent replication across studies. The schema has been applied and evaluated across ten case studies.

\subsubsection{Context Setup Module}
This module prepares the dynamic elements required for each simulation run, including avatar profiles, environment states, and initial memory.

\noindent\blue{\textbf{Avatar Profile Generation:}
Each case study targeted specific participant groups-such as older adults, students, or tech novices-to elicit feedback aligned with its research objectives. To simulate these populations, we generate avatar profiles using structured profile representations. These profiles include demographic attributes (e.g., age, occupation) and other relevant attributes (e.g., technology familiarity, household type) that are sampled based on a given distributions (e.g., the reported ones in the original studies). Each avatar is also assigned a TIPI-based personality score and a narrative description reflecting the intended characteristics (detailed in system design).
}
The assigned personality traits, combined with the demographic and other relevant attributes, are used to generate detailed background narratives for each avatar—such as lifestyle routines, communication styles, and preferences—to support lifelike behavior during interaction (see Appendix~\ref{appendix: context_setup} for persona narrative examples).

\noindent\textbf{Environment Configuration:} The physical environment is initialized to reflect the study setting (e.g., home, office) and \blue{by default uses a one-bedroom smart home layout}. Rather than including all visible objects, the system selects only those likely to emerge during interaction \blue{- including appliances (e.g., floor lamp, air conditioner, fan), tools (e.g., vacuum, remote control), and furniture (e.g., sofa, table)-}supporting clarity and task relevance. These curated interactive objects are accessible to both assistant agent and avatar and serve as cues for context-aware interactions. \blue{Researchers can optionally reconfigure the space to simulate different living arrangements, such as shared-living environments or single-family houses, depending on study requirements. This flexibility allows case studies to reflect diverse domestic contexts, which may influence interaction patterns} (see Appendix~\ref{appendix: context_setup} for an environment example). 

\noindent\textbf{Memory Management:} All contextual elements—including avatar profiles and environment configurations—are initialized during the first iteration and reloaded in each subsequent simulation round. This supports coherent, persona-aligned responses during assistant agent and avatar interactions. Since all case studies are conducted within a short time window (typically within a few hours), long-term memory management is not implemented.

\subsubsection{Assistant Agent--Avatar Interaction Module}
This module governs the runtime execution of the simulation by managing the dialogue and decision-making processes between the assistant agent and avatars.

\noindent\textbf{Interaction Dynamics:} Both assistant agent and avatars are powered by GPT-4o. They operate independently, generating responses based on their individual prompts, task goals, personas, and accessible context (see Appendix~\ref{appendix: aaa_interaction}). Interaction histories, memory states, and environmental changes are systematically logged to support traceable analysis and reproducibility.

\noindent\textbf{Knowledge Isolation:} The assistant agent and avatars do not share internal states or reasoning processes. The assistant agent cannot access the avatar's internal state unless made explicit during interaction. For example, the assistant agent cannot reference or infer the avatar’s unspoken intentions or internal states unless revealed during the conversation. 
Because these interactions are generated dynamically using large language models, we implement strategies to reduce factual inconsistencies or hallucinations. Each response is constrained by (1) Memory snapshots that include shared interaction history and role-specific internal reasoning for the assistant agent and avatar, (2) Real-time environmental context derived from the simulation state, and (3) Role-specific knowledge boundaries that enforce strict separation between the assistant and avatar's accessible information.

More advanced techniques such as retrieval augmentation~\cite{lewis2020retrieval, guu2020retrieval} or factual consistency checking~\cite{kryscinski2019evaluating, ji2023survey} can be applied to further reduce potential hallucination. However, we did not observe significant hallucinations with our current strategy, so we chose not to incorporate them in this version.

\noindent\blue{\textbf{Scenario Modeling:}
Several case studies included example scenarios to guide participant interactions, such as sample dialogues, storyboard narratives, or descriptions of desired experimental scenes. When these scenarios were provided in textual form, they were directly incorporated into the simulation instructions. For non-textual formats, we translated the content into narrative-driven textual instructions. In particular, studies employing storyboards were converted into structured narratives that preserved the original interaction flow and conversations.}

\noindent\blue{\textbf{Assistant Agent and Avatar Behavior Policies:}
Both the assistant agent and the avatar were implemented using a large language model (LLM), enabling them to generate naturalistic, role-appropriate behaviors. Assistant agent behaviors were guided by its designs and task objectives described in the original user studies. Depending on the study scenario, the assistant agent was assigned distinct roles to match the requirements of each study scenario—for example, serving as proactive recommender, experiment facilitators, or context-aware interactor. The avatar, in turn, was configured to simulate a human participant—responding to assistant prompts, expressing preferences, making decisions, and reflecting on experiences as appropriate for the study context.}

\blue{Treating both the assistant agent and avatar as role-playing participants, we provided detailed, role-specific instructions outlining their expected behaviors and responsibilities. For studies involving multi-turn interactions, both the assistant agent and avatar were equipped with conversation memory to retain prior exchanges and ensure interaction coherence. The assistant agent was configured to support either single-turn or multi-turn dialogue, depending on the study's requirements. The number of interaction rounds was customizable to replicate the original experimental protocol. These behavior policies governed when and how the assistant initiated actions, provided suggestions, or responded to avatar input in a manner consistent with the study's design.}

\begin{figure}[t]
    \centering
    \includegraphics[width=1\linewidth]{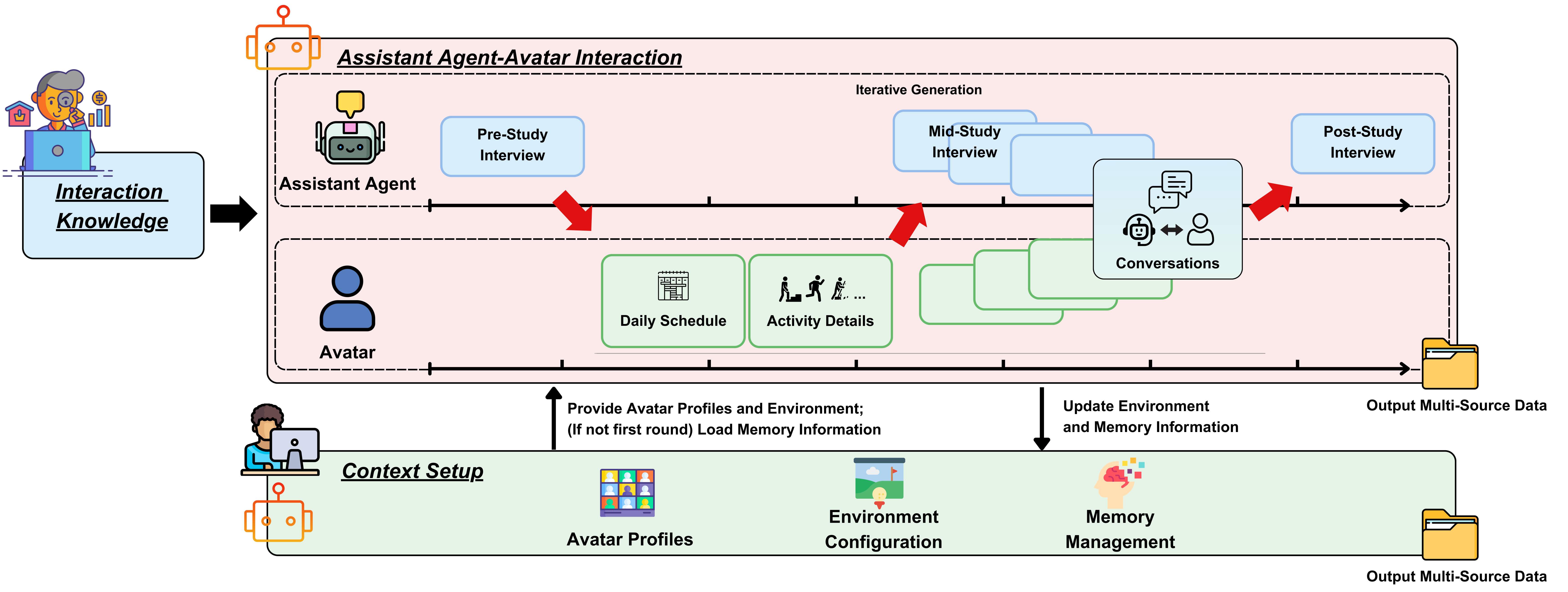}
    \vspace{-20pt}
    \caption{\revise{Simulation workflow of \system{}. In the initialization phase, researchers provide interaction knowledge, avatar profiles, environment configurations, and memory setup. During runtime, the Assistant Agent–Avatar Interaction module executes iterative study rounds, generating dialogues, updating memory and environment states, and producing simulation results.}}
    \label{fig:workflow}
    %\vspace{-15pt}
\end{figure}

\subsection{Simulation Workflow}

Fig.~\ref{fig:workflow} illustrates the execution flow of \system{}.  
The simulation proceeds through a structured sequence of phases designed to support realistic, traceable human--assistant interactions. Below, we describe the key steps that take place during each simulation run. A recorded demo based on the platform was also submitted.

\textbf{1. Initialization Phase.} The system begins by loading the interaction knowledge files and configuring the simulation context. This includes avatar profiles, environment layout, and prior event history to ensure consistency and completeness. %A sample interaction knowledge file is provided in Appendix~\ref{appendix:interaction_knowledge}.

% The simulation begins with an initialization phase, where the system loads the interaction knowledge files and configures all context setup—including avatar profiles, environment layouts, and event histories—to ensure completeness and consistency A representative example of the interaction knowledge files is provided in Appendix~\ref{appendix:interaction_knowledge}..  

\textbf{2. Pre- and Mid- Study Interviews (Optional).} If defined by the study design, the assistant conducts a pre-study or mid-study interview with the avatar. The avatar’s responses are logged and later used for post-simulation evaluation and comparison.

% At the beginning of each simulation round, if defined by the researcher, a pre-study interview phase is conducted.   
% These responses are recorded and used as reference points for post-simulation evaluation.
% Each simulation iteration consists of three sequential components: basic schedule generation, detailed activity enrichment, and assistant–avatar conversation interaction. 

\textbf{3. Schedule Generation.} Human--assistant interactions are often grounded in daily routines. The system generates a basic, time-stamped daily schedule for each avatar using a GPT-4o prompt (see Appendix~\ref{appendix:activity_generation}). The prompt is conditioned on the avatar's profile, environment, and memory state. Typical activities might include breakfast preparation, checking notifications, or starting work.

\textbf{4. Activity Enrichment.} Each scheduled activity is expanded with additional context using another prompt (see Appendix~\ref{appendix:activity_generation}).
The enrichment includes motivations, subtasks, expected environmental interactions, and potential challenges. 
This enrichment process supports fine-grained avatar behavior modeling and provides the assistant agent with context for initiating targeted interactions, enabling more realistic interaction details and capturing potential opportunities for timely or context-relevant interaction.  

% Human–assistant interactions, whether in real-world case studies or in simulated environments, are typically initiated or triggered by human activities.  
% Following this, the system generates a basic daily schedule for the avatar using a prompt-based approach with GPT-4o, conditioned on the avatar’s profile, environmental context, and current memory state(see Appendix~\ref{appendix: daily_schedule} for the daily schedule generation prompt).

% This schedule includes time-stamped activities structured to reflect realistic daily routines, such as preparing breakfast, checking notifications, or completing a work-related task.  
% These base-level actions establish a temporal backbone for the simulation, structuring when and where interactions are likely to occur.
% For each basic schedule, the system generates activity details descriptions that expand the activities with motivations, subtasks, expected environmental interactions, and potential challenges (see Appendix~\ref{appendix: activity_details} for the activity details generation prompt).

\textit{Example:} For a base activity such as ``Alex starts working on a new art project at the table in the main room,'' the system generates a detailed scenario in which Alex draws inspiration from a recent trip and adjusts the lighting to enhance his workspace. This elaboration reveals his motivation (creative expression inspired by ocean memories), subtasks (preparing paints, adjusting blinds, evaluating progress), environmental interactions (engaging with sunlight, paint tools, hydration), and potential challenges (distraction, time unawareness).

% . For example, given a base schedule such as ``Alex starts working on a new art project at the table in the main room'', the system generates a detailed scenario in which Alex draws inspiration from a recent trip and adjusts the lighting to enhance his workspace. This elaboration reveals his motivation (creative expression inspired by ocean memories), subtasks (preparing paints, adjusting blinds, evaluating progress), environmental interactions (engaging with sunlight, paint tools, hydration), and potential challenges (distraction from unrelated thoughts, time unawareness).

% This process enhances behavioral realism and provides the assistant with fine-grained cues for initiating targeted interventions.

\textbf{5. Assistant Agent--Avatar Interaction.} Based on the enriched context, the assistant may engage with the avatar either proactively or responsively, depending on the scenario logic. Both roles are powered by separate GPT-4o instances (see examples in Appendix~\ref{appendix: aaa_interaction}). They operate independently, with shared memory (e.g., conversation history) and non-shared attributes (e.g., goals, role-based knowledge). This asymmetry mimics real-world facilitator–participant dynamics.

% The assistant may respond to or proactively intervene in these activities depending on the study design—such as offering reminders, giving suggestions, or conducting mid-study interviews.

% Each assistant–avatar interaction is grounded in current environmental cues, memory context, and the assistant’s scoped task goals.  
% During the interaction phase, the assistant agent and avatar roles are powered by separate LLM instances, generating multi-turn dialogues that reflect the scenario logic defined in the interaction knowledge configuration.  
% The two roles operate with both shared (e.g., conversational history) and non-shared memory structures, including distinct avatar profiles, experimental roles, and assigned tasks.  
% This design simulates the asymmetry of knowledge typically seen between human experiment facilitators and participants.  
% Interactions may consist of brief question–answer exchanges or longer reflective interviews, depending on the study objectives.

% Each simulation iteration consists of three sequential components: basic schedule generation, detailed activity enrichment, and assistant–avatar conversation interaction.  

\textbf{6. State Update.} After each interaction, the system updates the avatar's memory and the environmental state to reflect what happened. These updates carry over into the next simulation round, allowing the simulation to maintain continuity and coherence.

% After each iteration, the system updates both the avatar's memory and the environmental state to reflect the actions taken and dialogue outcomes.  
% These updates enable the next iteration to build on past behaviors and maintain internal coherence across the simulation timeline.

\textbf{7. Post-Study Interview.} At the end of the simulation, the assistant may conduct a post-study interview to assess avatar’s perspectives and the assistant agent’s performance. This mirrors real-world methods used in evaluations of human–assistant interaction research.

\textbf{8. Logging and Output.} All simulation output—including multi-source data such as schedules, activity details, dialogue transcripts, environmental states, and memory logs—are saved in structured text formats. This supports reproducibility and facilitates analysis.

\section{System Evaluation}
\label{sec:casestudies}

\blue{\subsection{Case Study Selection}}

% Explain the motivation for replication-based evaluation
% Describe criteria for selecting the 10 case studies

\blue{\textbf{Selection Criteria:}} Evaluating the realism of human decision-making in human–assistant interactions remains a challenge due to the limited availability of publicly accessible datasets that capture both observable behaviors and internal contextual states. We assess \system{}'s effectiveness by examining whether it can replicate the key findings or observations from prior human-participated studies, i.e., a replication-based evaluation strategy.
To find relevant studies, we began by surveying the HCI literature from major venues such as IMWUT and CHI  using the keywords ``voice assistant,'' ``conversational agent,'' and ``proactive system.'' 
The studies were selected following these criteria:

\begin{itemize}
    \item The interaction must involve human participants and an assistant agent executing a clearly defined task. The assistant agent's behavior must remain consistent across all participants and may be instantiated via a system prototype or a human confederate in a Wizard-of-Oz configuration.
    
    \item The studies must be well-documented for possible replication. All interaction topics or tasks must be accessible to participants without requiring specialized domain knowledge.
    
    %\item The study must not target special populations (e.g., individuals with disabilities), as such groups are typically underrepresented in the datasets used to train generative AI models. Consequently, the simulator may lack sufficient background knowledge to model these populations reliably.
    
    \item The interaction must be conducted via or convertible to text, as the current version of the simulation environment supports only textual modalities for both input and output.

    \item The case studies reflect diverse research focus or themes on human-assistant interactions.
\end{itemize}

Following these criteria, we identified ten case studies~\cite{zargham_i_2022, clark_what_2019, chin2024like, dubiel2024hey, wei2021understanding, zargham_understanding_2022, cha_hello_2020, reicherts_may_2021, oh_better_2024, liu_understanding_2023} that reflect four key themes in assistant research, including \textit{Personalization and Social Framing}, \textit{Proactivity and Context-Awareness}, \textit{Managing Attention and Interruptibility}, and \textit{User Control and In-Situ Configuration}. Table~\ref{tab:cs_intro} provides an overview of their research objectives, study method, participant and physical setups, and evaluation metrics. 
%\textit{CS1--CS3 are adapted from~\cite{zargham_i_2022, clark_what_2019, chin2024like}; CS4--CS6 are adapted from~\cite{dubiel2024hey, wei2021understanding, zargham_understanding_2022}; CS7--CS8 are adapted from~\cite{cha_hello_2020, reicherts_may_2021}; and CS9--CS10 are adapted from~\cite{oh_better_2024, liu_understanding_2023}.}

\begin{table}[!p]
    \caption{Summary of case studies, research objectives, methods, setup, and evaluation metrics.}
    \vspace{-10pt}
    \includegraphics[width=0.95\linewidth]{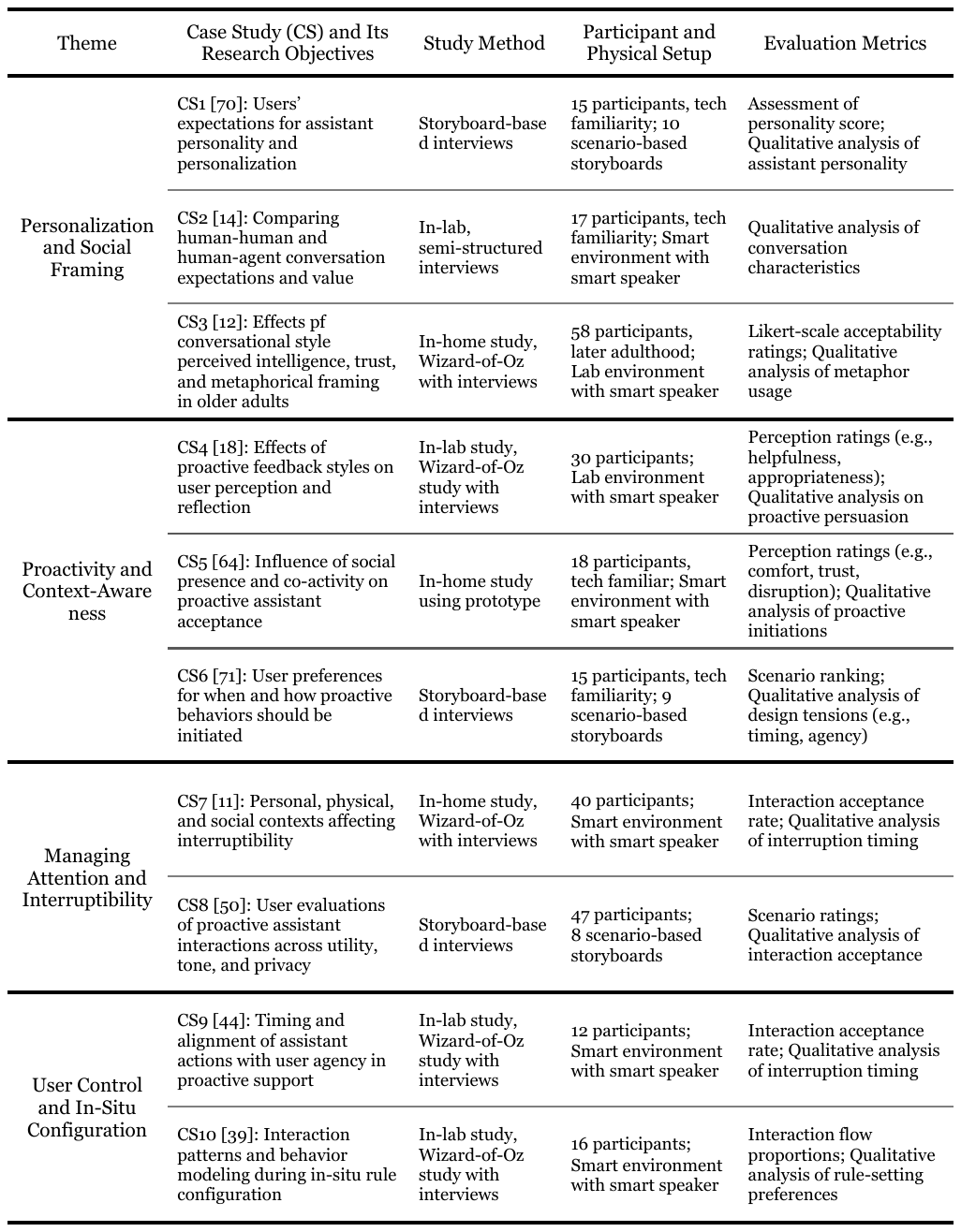}
    \label{tab:cs_intro}
\end{table}

\blue{\subsection{Evaluation Methods}}
\label{sec:evaluation_methods}
\textbf{We define a successful replication as the ability of \system{} to generate simulated outcomes that align with the core research questions and findings of each original study.} Rather than attempting to reproduce participant responses verbatim, we assess whether the simulation captures comparable interaction patterns and assistant-driven effects. \blue{We evaluate \system{}'s effectiveness through a replication-based approach that assesses whether simulated interactions can reproduce key findings from prior human-subject studies. Our evaluation employs two complementary methods designed to capture both high-level semantic alignment and detailed behavioral consistency across diverse study designs.}

\subsubsection{Semantic Similarity to Research Question Answers:} 
For each study, we identify the formally stated research questions and summarize the key findings. Simulated answers are generated from assistant \blue{agent--avatar interactions, avatars' feedback ratings and surveys},  and their alignment with the original findings is assessed using semantic similarity metrics based on embedding models. This approach evaluates whether the simulator yields insights that are conceptually consistent with those derived from human participants.

\textbf{Data Processing and Analysis Pipeline:}
\label{sec:memorize}
\begin{itemize}
    \item \textbf{Response Extraction:} We systematically extracted response data from two distinct sources: original study findings and \system{}-generated outputs. For original findings, we identified the research questions formulated in each paper and extracted corresponding key findings from relevant sections, leveraging the explicit mappings between research questions and results typically provided by the authors. For \system{}-generated data, we collected multi-turn interaction logs between assistant agents and avatars, along with completed questionnaires that paralleled the original evaluation instruments.
        
    \item \textbf{Summarization and Embedding:} 
    Both data sources underwent identical summarization procedures using GPT-4o with uniform prompt structures designed to generate focused summaries that directly address each research question in a consistent question-and-answer format. \revise{Each summarization request to GPT-4o was executed as an independent API call without shared conversational history, ensuring no cross-contamination between data sources.} We then used the all-mpnet-base-v2 model from Sentence-Transformers~\cite{reimers2020making, song2020mpnet} to generate embedding representations of the summarized responses. Semantic similarity was quantified by computing cosine similarity between embedding vectors for each research question pair. \textit{Structured prompts and generated summary examples are provided in Appendix \ref{appendix: summary_revise_prompt} and Appendix \ref{appendix: summary}.}

\end{itemize}

\textbf{Bias Mitigation and Human Validation:}
\begin{itemize}
    \item \textbf{Bias Mitigation Strategies:} We implemented several safeguards to minimize evaluation bias: (1) bottom-up pattern emergence, where avatars follow detailed activity sequences and interaction protocols without direct exposure to high-level study hypotheses, allowing behavioral patterns to emerge naturally rather than being artificially guided toward expected outcomes; (2) content abstraction using high-level findings rather than verbatim quotes, where we extracted conceptual themes and synthesized insights from both original studies and simulated outputs instead of relying on direct textual matches, reducing the risk of surface-level similarity inflating semantic alignment scores; and (3) summary-level comparison of synthesized insights rather than raw text, where we used structured summarization prompts to generate comparable abstractions of key findings from both sources, ensuring that evaluation focused on conceptual alignment rather than linguistic similarity, thereby avoiding bias from writing style differences or terminology variations between original papers and simulation outputs.

    \item \textbf{Manual Summary Quality Check:} To ensure the reliability of our automated similarity scoring, two researchers independently reviewed all generated summaries to verify that key themes and findings were preserved during the summarization process. We implemented systematic quality control measures including cross-validation between reviewers and structured assessment criteria for thematic completeness and conceptual accuracy. To control for consistency in summarization style, we designed structured prompts that explicitly instructed GPT-4o to generate summaries at a consistent level of abstraction, balancing high-level conceptual insights with specific scenario-based evidence. This prompt-guided approach ensured that both original study summaries and \system{} simulation summaries maintained comparable depth and focus, minimizing variability in the level of detail that could affect similarity comparisons. Any discrepancies in summary quality or consistency were resolved through researcher consensus and iterative refinement of the summarization process.
\end{itemize}

\subsubsection{Interaction Log Analysis:} 
To contextualize the semantic similarity results, we analyze assistant agent–avatar interaction logs using a mixed-methods approach, combining quantitative metrics with qualitative interpretation \blue{to assess whether \system{} reproduces the behavioral patterns observed in the original studies. Where possible, we replicated the original analytical procedures to ensure comparability.}

\blue{\textbf{Data Processing and Analysis Pipeline:}}
\begin{itemize}
    \item \textbf{Quantitative patterns:} \blue{Across the 10 selected case studies,} we conduct\blue{ed} statistical analysis \blue{encompassing response rates, interaction timing patterns, preference rankings, and Likert-scale rating distributions} to evaluate behavioral metrics. \blue{Given that detailed conversational and personal data are typically not released due to privacy constraints, we directly compared calculated statistics that align with those reported in the original studies.} 

    \item \textbf{Qualitative evidence:} \blue{Verbatim quotes and trace logs serve as strong empirical evidence for supporting theoretical claims~\cite{moravcsik2014transparency}. Given the diversity in human perspectives, we identify recurring thematic patterns in conversational responses and viewpoints, which reflect broader collective trends across participant groups~\cite{flyvbjerg2006five}.} Representative dialogues and reasoning traces are extracted to illustrate parallels with the original study's reported behaviors. \blue{These patterns are then compared with participant quotes and behavioral descriptions from the original studies to assess alignment.}
    
\end{itemize}

\textbf{Bias Mitigation and Human Validation:}
\begin{itemize}
    
    \item \textbf{Quote Selection Process:} Representative quotes were selected through a systematic process where two researchers independently identified exemplar dialogues that best illustrated the quantitative patterns observed in each study. Disagreements in quote selection were resolved through structured discussion and consensus-building. 
    
    \item \textbf{Manual Review of Quantitative Results:} Human reviewers verified that the selected measures and visualizations matched those used or implied in the original studies. When original procedures were unclear, reviewers adapted the analysis to reflect the reported behavioral trends and ensured interpretive consistency.

\end{itemize}

Together, these methods allow us to evaluate the fidelity of simulation outcomes in terms of both high-level semantic alignment and detailed behavioral consistency. \blue{We present the results using these methods in Sections \ref{sec: aggregate analysis} and \ref{sec: case analysis} for evaluation for analysis and in Section \ref{sec:cross-model} for validation.} This framework provides a robust and interpretable basis for comparing simulated and empirical results across diverse study designs and reporting formats. 

\subsection{Aggregate Replication Accuracy Across Case Studies}
\label{sec: aggregate analysis}

This part presents the replication accuracy across 10 case studies.
To fairly compare the semantic similarity between the original study and generated simulated results on each research questions, we summarized the extracted original findings and the simulated interaction data and questionnaire feedback for each RQ.

\begin{figure}[t]
    \centering
    \includegraphics[width=1\linewidth]{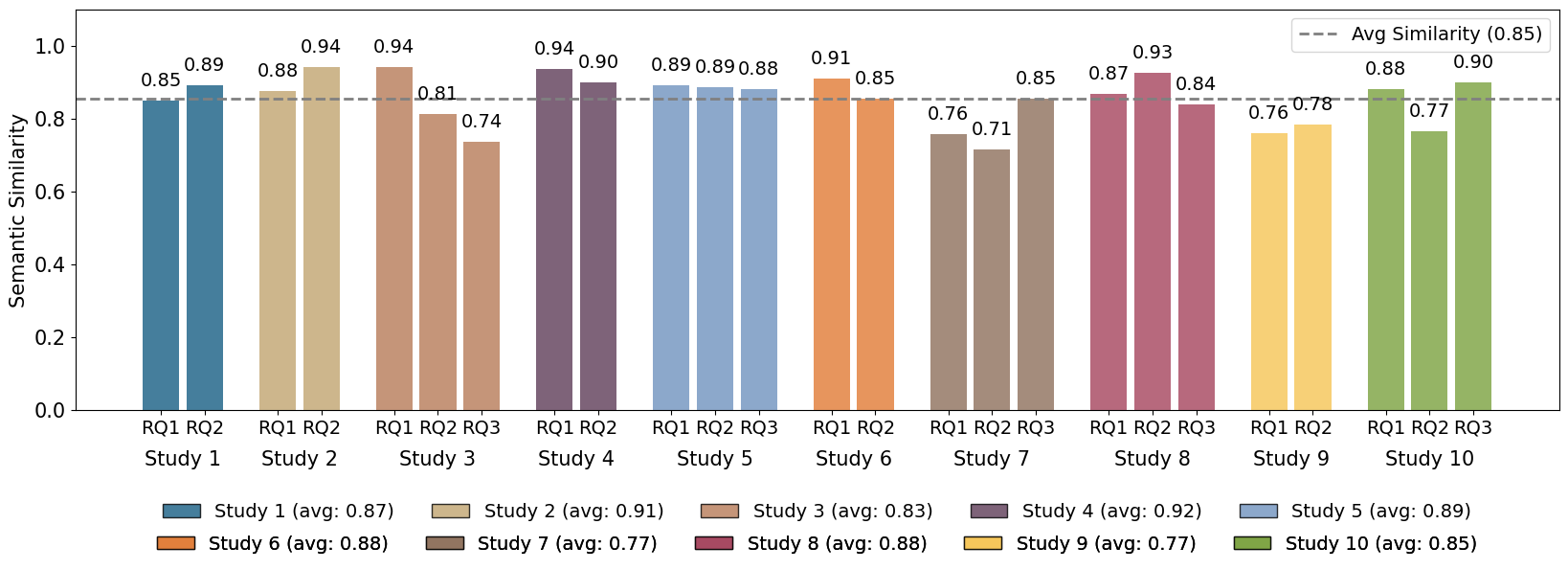}
    \vspace{-20pt}
    \caption{Semantic similarity scores comparing simulated and original study responses for each research question (RQ) across 10 case studies. The red line shows the overall average similarity (0.85). Study-specific average similarities are in the legend.}
    \Description{}
    \label{Figure:rq_sim}
\end{figure}

\begin{figure}[t]
    \centering
    \begin{subfigure}[H]{0.45\textwidth}
        \centering
        \includegraphics[width=\linewidth]{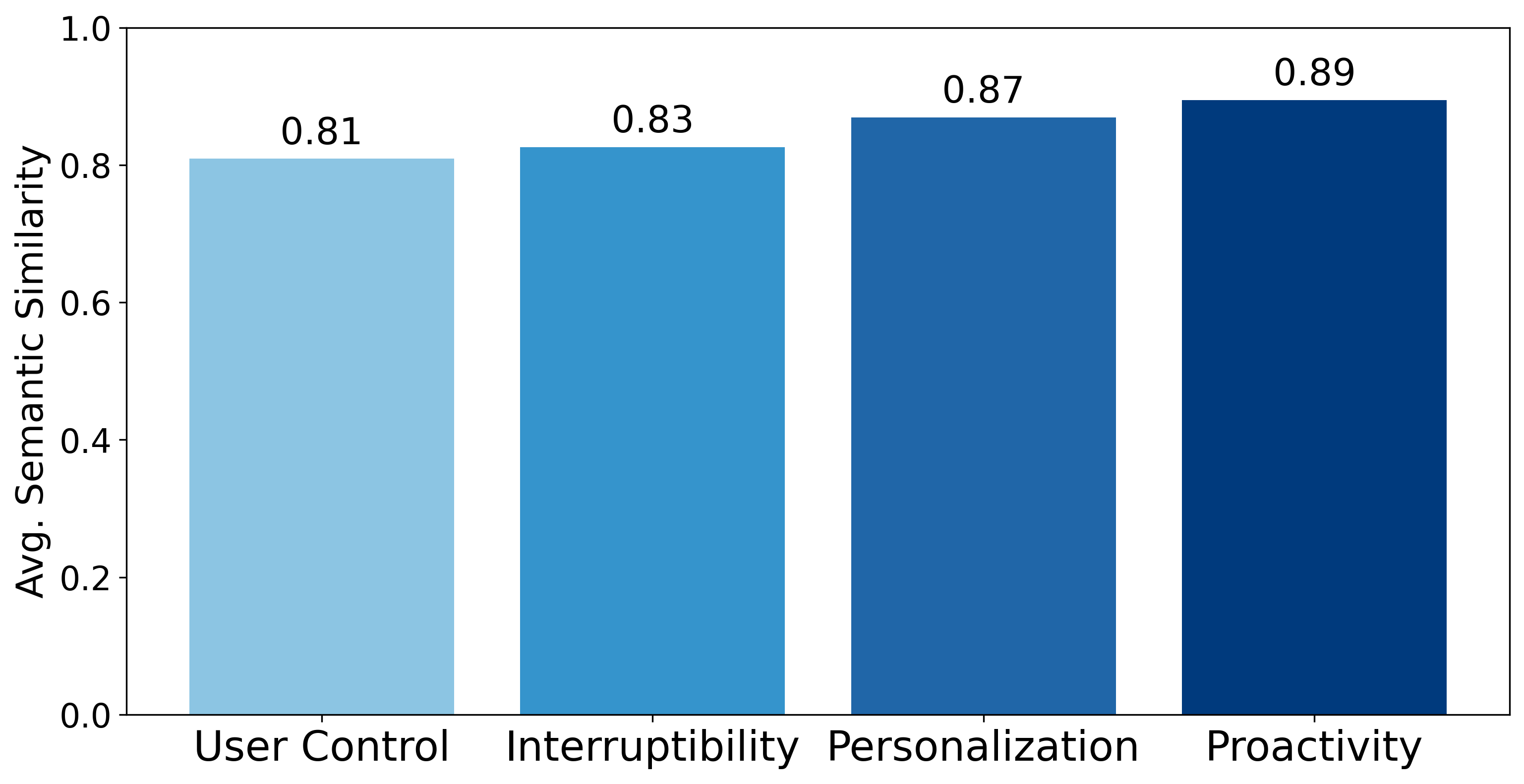}
        \caption{Average semantic similarity grouped by study theme.}
        \label{Figure:theme_sim}
    \end{subfigure}
    \hspace{0.03\textwidth}
    \begin{subfigure}[H]{0.45\textwidth}
        \centering
        \includegraphics[width=\linewidth]{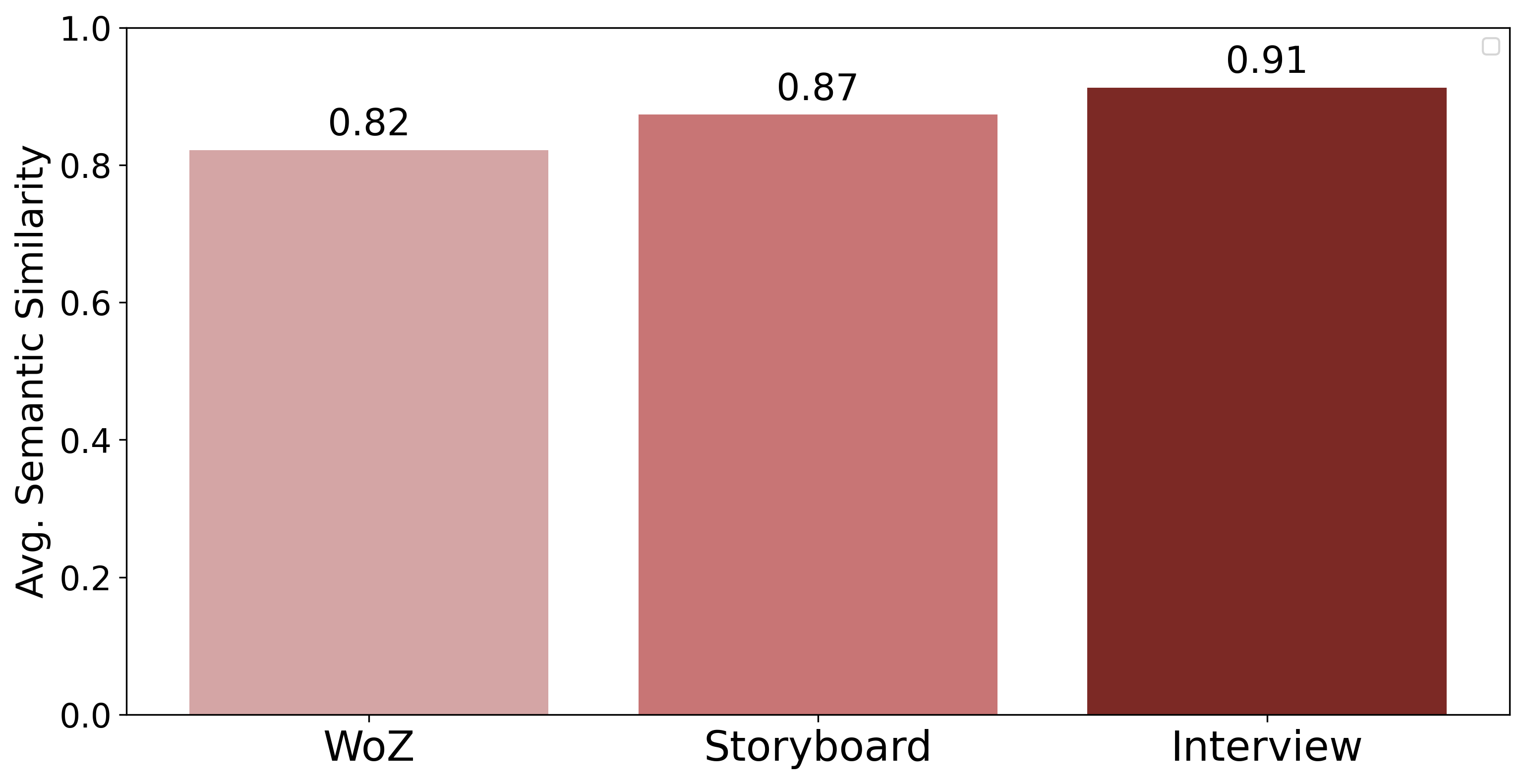}
        \caption{Average semantic similarity grouped by study mode.}
        \label{Figure:mode_sim}
    \end{subfigure}
    \vspace{-10pt}
    \caption{Average semantic similarity scores grouped by (a) study theme—User Control, Interruptibility, Personalization, and Proactivity—and (b) study mode—Wizard-of-Oz (WoZ), Storyboard, and Interview.}
    \Description{}
    \label{Figure:group_sim}
\end{figure}

\iffalse
\begin{figure}[t]
    \centering
    \begin{subfigure}[b]{0.53\textwidth}
        \centering
        \includegraphics[width=\linewidth]{figure/Smart Home.pdf}
        \caption{The average semantic similarity score is 0.88. Cell color encodes similarity level: scores from 0.70–0.79 are shown in \textcolor{Salmon}{red}, 0.80–0.89 in \textcolor{Dandelion}{yellow}, and 0.90–1.00 in \textcolor{ForestGreen}{green}.}
        \label{Figure:rq_sim}
    \end{subfigure}
    \hspace{0.03\textwidth}
    \begin{subfigure}[b]{0.42\textwidth}
        \centering
        \includegraphics[width=\linewidth]{figure/group.png}
        \caption{Average semantic similarity scores grouped by study theme (top) and study mode (bottom).}
        \label{Figure:theme_sim}
    \end{subfigure}
    \caption{Semantic similarity analysis across simulated and original study findings. (a) shows RQ-level scores per study. (b) summarizes theme- and mode-level averages.}
    \label{Figure:group_sim}
\end{figure}
\fi

Fig.~\ref{Figure:rq_sim} visualizes the distribution of similarity scores across all 25 RQs. Each bar represents the similarity score for a single RQ, grouped and color-coded by case study. The average similarity per case study is also shown in the legend, offering a study-level view of replication quality. \blue{We further grouped the results by (a) study theme and (b) experimental mode to visualize and analyze the replication quality of \system{}}. 
As shown in Fig.~\ref{Figure:theme_sim}, studies focused on \textit{Proactivity} and \textit{Personalization} achieved the highest semantic alignment (0.89 and 0.87, respectively), while \textit{User Control} studies yielded the lowest average similarity (0.81). We observed that study theme was associated with differences in semantic similarity, prompting further analysis based on study mode to explore potential underlying causes. Notably, both User Control studies employed a Wizard-of-Oz (WoZ) methodology, which often introduces variability due to dynamic and spontaneous human activities. As a result, reactions and responses triggered by activities only partially aligned with the original case study findings. Moreover, the original case studies did not provide detailed behavioral data, limiting our ability to adjust the simulation to better match the original conditions.
We further analyzed the results by study mode. 
As shown in Fig.~\ref{Figure:mode_sim}, simulations of \textit{Interview}-based studies achieved the highest average similarity (0.91), followed by \textit{Storyboard} (0.87), and \textit{WoZ} (0.82). Interview and storyboard studies typically provide structured questionnaires or detailed scenario examples, offering clear rationales for user responses and reducing ambiguity during simulation. This explains why they achieve higher semantic similarity scores compared to WoZ-based studies. Overall, these findings confirm that our simulation framework achieves a high semantic similarity in reproducing study findings, with variations closely tied to study design and some misalignment occurring when more spontaneous factors are involved.

\subsection{Case Study-Level Evaluation Results}
\label{sec: case analysis}

To better interpret \system{}'s effectiveness across diverse research contexts, we conducted a study-level analysis based on the full conversational transcripts between the assistant agent and avatar agents. For each case study, we report both qualitative observations and quantitative patterns extracted from the interaction logs,highlighting specific replication outcomes and discussing unique factors that influenced simulation quality.
Before presenting case-by-case results, we summarize several key insights derived from the ten replicated studies. Each insight is supported by specific examples, which are elaborated in the corresponding case study sections below \blue{and in Appendix \ref{appendix: case_studies}}. %While these findings demonstrate strong replication capabilities, we also observed several important limitations, which will be elaborated in Section~\ref{sec:discussion_limitation}. 

\begin{itemize}
    \item \textbf{Preservation of High-Level Behavioral Patterns}: 
    \system{} preserved core behavioral patterns observed in original human–assistant interaction experiments across the ten replicated studies, such as activity structures, conversational framing, proactive interaction timing, and smart home rule programming strategies. %~\ref{sec:cs1}

    \item \textbf{Alignment of Thematic Reasoning}:  
    \system{} avatars demonstrated reasoning that paralleled human participants.  
    Both quantitative outcomes (e.g., conversation response rates, scenario rankings) and qualitative rationales (e.g., desire for autonomy, sensitivity to timing, comfort with emotional tone) showed strong thematic alignment, making it suitable for simulating user-centered interaction studies.

    \item \textbf{Context-Sensitive Interaction Modeling}: 
    \system{} reflected users' tendency to adjust engagement based on contextual factors such as physical activity state, emotional readiness, and privacy concerns, reinforcing its utility for studying experiments that include contextual information as a critical factor.

    \item \textbf{Strong Modeling of Rule-Based Configuration Behavior}:  
    In the in-situ programming studies, \system{} mirrored users’ programming approaches, such as starting with direct control commands and progressively refining rules based on spatial or environmental cues, demonstrating its ability to replicate smart home interaction and collaboration-related studies.

    \item \textbf{Potential for Group-Level Reflection Simulation}:  
    Our evaluation suggests that \system{} is capable of generating human-like reflections across diverse avatars, making it promising for applications that require capturing group-level opinions, participatory design feedback, or early-stage concept evaluations.

    %\item \textbf{Utility as a Pre-Deployment Testbed for Assistant Models}  
    %\system{} provides a scalable platform for evaluating intelligent assistant behaviors and gathering preliminary user feedback before conducting resource-intensive real-world participant studies.
\end{itemize}

%%%%%%%%%%%%%%%%%%%%%%%%%%%%%%%%%%%%%%%%%
% GROUP 1
%%%%%%%%%%%%%%%%%%%%%%%%%%%%%%%%%%%%%%%%%

\subsubsection{Theme: Personalization and Social Framing}
\label{sec: 5.5.1}
%\textcolor{red}{Add intro sentences about why these studies are grouped together. What similarities they share in discussing personalization and social framing. Discussion: Current agent designs are limited by exisiting technology. The studies assign agents with more human-like traits-TIPI scoring, conversation style and tone-using storyboard/woz to investigate user acceptance to a "personalized" agent.}

This group of studies investigates how personalization and social framing influence user perceptions of home assistants. Despite differences in focus, all three studies assign agents human-like traits—such as personality profiles (e.g., TIPI scoring), conversational style, and tone. They tend to use storyboard or Wizard-of-Oz (WoZ) methods to explore user acceptance of more socially attuned, personalized agents, due to current technological limitations in prototype design. \blue{We present Case Study 1 below, with Case Studies 2 and 3 detailed in Appendix \ref{appendix: theme1}.}

%This set of studies explores how users relate to human assistants through the lens of personalization, conversational style, and perceived social intelligence. The three studies—\textit{Personalization Preferences}~\cite{zargham_i_2022}, \textit{Conversational Qualities in Dialogue}~\cite{clark_what_2019}, and \textit{Conversational Style for Older Adults}~\cite{chin2024like}—emphasize the social aspects of interaction design, including customization preferences, dialogue quality, and the metaphorical framing of agent personas.
\begin{figure}[t]
    \centering
    \begin{subfigure}[t]{0.45\textwidth}
        \centering
        \includegraphics[width=\linewidth]{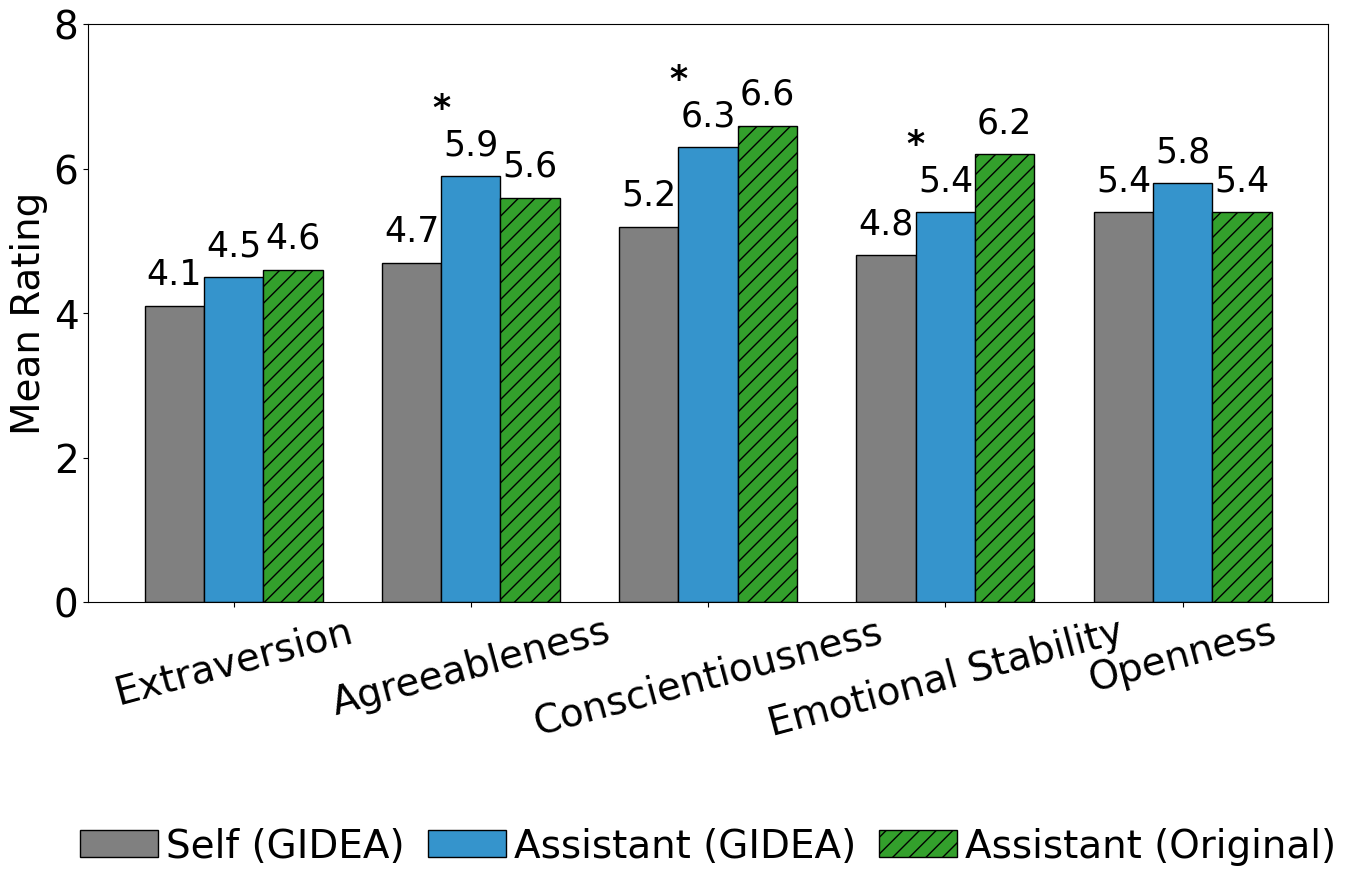}
        \caption{Case Study 1: Mean personality trait ratings comparing users' self-perception and their imagined assistant (agent), across simulated and original datasets.
        * Significant difference between Self and Assistant ratings in GIDEA (p < 0.01)}
        \label{Figure:cs1}
    \end{subfigure}
    \hspace{0.03\textwidth}
  \begin{subfigure}[t]{0.45\textwidth}
    \centering
    \renewcommand{\arraystretch}{0.9}
    \includegraphics[width=\linewidth]{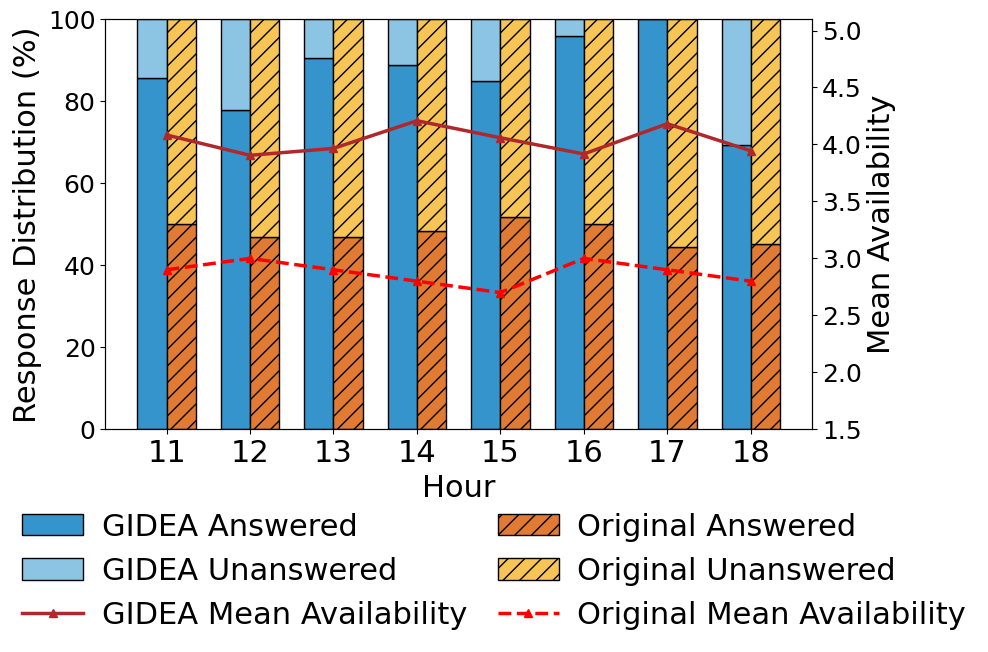}
    \caption{Case Study 5: Hourly response distributions and mean availability between GIDEA and the original user studies.}
    \label{Figure:cs7}
  \end{subfigure}
    \caption{Personalization and Social Framing.}
    \label{Figure:cs110}
\end{figure}

\begin{table}[t]
\captionsetup{skip=2pt}
\caption{Case Study 1: Aligned Perceptions from Simulated Avatars and Human Participants}
\small
\renewcommand{\arraystretch}{1.3}
\begin{tabularx}{\textwidth}{@{}>{\centering\arraybackslash}p{3.3cm} >{\raggedright\arraybackslash}X >{\raggedright\arraybackslash}X@{}}
\toprule
\textbf{Theme} & \multicolumn{1}{c}{\textbf{Simulated Avatar Quote}} & \multicolumn{1}{c}{\textbf{Human Participant Quote}} \\
\midrule
\textbf{Emotionally Adaptive Response} & 
``It felt like the assistant was genuinely attuned to my \textbf{preferences} and emotions, which was comforting.” & 
``It should adjust the information content based on my \textbf{desires}, not necessarily behave like me” 
\\
\arrayrulecolor{black!30}\midrule
\textbf{Risk of Misinterpretation} & 
``Misinterpreting mood could lead to 
\textbf{frustration}.” & 
``If it misjudges my mood, it would work \textbf{horribly wrong}.” \\
\arrayrulecolor{black!30}\midrule
\arrayrulecolor{black}
\textbf{Anthropomorphism Boundaries} & 
``I do not want the assistant to feel \textbf{too human}. It should still feel like a tool I can manage.” & 
``I don’t need a piece of software to show me \textbf{empathy}, I know it’s programmed” \\
\bottomrule
\end{tabularx}
\label{tab:study1_matched_quotes}
\end{table}

\begin{itemize}
\item \textbf{Case Study 1 - Personalization Preferences}\label{sec:cs1}
%How do users imagine the personality of their desired home assistant?  
%In which ways do users want to customize and personalize their home assistants?

\textbf{- Overview.}
Zargham et al.\cite{zargham_i_2022} explored how participants imagine the ideal personality and customization features of home assistants using a storyboard-based experimental method. To replicate this case study, we initialized simulated avatars in \system{} with the same mean personality trait scores as the original participants' self-assessments and exposed them to the same storyboard scenarios depicting daily interactions between home assistants and users.
% Based on the following analysis, \system{} demonstrates the ability to simulate human participants by preserving self-consistency in trait expression and reproducing key behavioral patterns observed in real-world studies. In the context of personality modeling, it captures both the baseline self-assessment tendencies and the relative shifts participants make when imagining ideal home assistants, highlighting its effectiveness in mirroring complex human judgment and adaptation processes.

\textbf{- Analysis.}  
As shown in Fig.~\ref{Figure:cs1}, gray bars show self-assessed personality traits, shared by both original participants and simulated avatars. Blue bars represent imagined assistant traits rated by simulated avatars in \system{}, and green bars reflect ratings from original participants.
The figure reveals a consistent trend: participants tended to envision assistants as having higher levels of agreeableness, conscientiousness, and emotional stability than themselves. This pattern appears in both the original study and the simulation. In particular, imagined assistants received significantly higher ratings than participants' self-assessments in agreeableness (\textit{t}(14) = -4.58, \textit{p} = .0004), conscientiousness (\textit{t}(14) = -4.43, \textit{p} = .0006), and emotional stability (\textit{t}(14) = -3.15, \textit{p} = .007). These findings support that participants idealize assistants as emotionally stable, supportive, and reliable.
Simulated avatars in \system{} also reflected participants' nuanced expectations for assistant behavior and interaction style. As shown in Table~\ref{tab:study1_matched_quotes}, the avatars echoed preferences for emotionally intelligent and adaptive communication—valuing assistants that respond sensitively to user mood and context. While this emotional responsiveness enhanced perceptions of supportiveness, both groups noted risks of misinterpretation. Inaccurate mood detection was seen as a potential source of discomfort or frustration. These considerations also shaped participants' views on anthropomorphism: both real and simulated participants preferred assistants that were not overly human-like, maintaining a clear distinction between a helpful tool and a human companion.
These findings show that \system{} has the ability to simulate human participants by preserving self-consistency in trait expression and reproducing key behavioral patterns observed in real-world studies. In the context of personality modeling, it captures both the baseline self-assessment tendencies and the relative shifts participants make when imagining ideal home assistants, highlighting its effectiveness in mirroring complex human judgment and adaptation processes.

\end{itemize}

%%%%%%%%%%%%%%%%%%%%%%%%%%%%%%%%%%%%%%%%%
% GROUP 2
%%%%%%%%%%%%%%%%%%%%%%%%%%%%%%%%%%%%%%%%%
\subsubsection{Theme: Proactivity and Context-Awareness}
\label{sec: 5.5.2}
This group of case studies examines strategies for proactive conversation initiation, addressing the challenge that current agent designs often lack appropriate triggers to naturally start interactions with users. The studies investigate how to define valid initiation triggers and explore the use of multi-source data to support context-aware proactive communication. \blue{We present Case Study 5 below, with Case Studies 4 and 6 detailed in Appendix \ref{appendix: theme2}.}

\begin{itemize}
\begin{table}[t]
\captionsetup{skip=2pt}
\caption{Case Study 5: Proactive Initiation and Adaptive Feedback}
\small
\renewcommand{\arraystretch}{1.3}
\begin{tabularx}{\textwidth}{@{}>{\centering\arraybackslash}p{4.2cm} >{\raggedright\arraybackslash}X >{\raggedright\arraybackslash}X@{}}
\toprule
\textbf{Theme} & \multicolumn{1}{c}{\textbf{Simulated Avatar Quote}} & \multicolumn{1}{c}{\textbf{Human Participant Quote}} \\
\midrule

\textbf{Opt-in Utterance Starters} & 
``The assistant’s gentle opening made it feel \textbf{optional}, which helped me stay in control while still being open to suggestions.” & 
``I like the way the system starts the conversation... I can just say ‘no’ when it’s inconvenient.” \newline
``It gives me the \textbf{choice} to interact, so it performs more like humans.” \\
\arrayrulecolor{black!30}\midrule
\arrayrulecolor{black}
\textbf{Adaptive and Emotionally Responsive Feedback} & 
``I want the assistant to learn about my preferences and tailor suggestions \textbf{based on my routine habits and mood}.” & 
``After reporting a bad mood, participants expected dynamic follow-up like \textbf{mood-relevant suggestions}.” \\

\bottomrule
\end{tabularx}
\label{tab:study5_matched_quotes}
\end{table}

\item \textbf{Case Study 5 – Proactive Initiation and Adaptive Feedback}
%RQ: How do in-situ features affect the programming model and facilitate the programming process?

\textbf{- Overview.}
Wei et al.~\cite{wei2021understanding} conducted a three-week in-home deployment study to examine user perceptions of proactive smart speaker behavior using a Wizard-of-Oz setup. To replicate this study, avatars performed common home activities while the assistant agent acted as the smart speaker, initiating conversations at contextually appropriate moments to make the interactions feel natural and well-timed. 

\textbf{- Analysis.}
Fig.~\ref{Figure:cs7} shows a stacked bar and line chart comparing the hourly response distributions and mean availability between \system{} and the original case studies. In the stacked bars, the darker color areas represent the proportion of responses that were answered within each hourly time period. We observe that the \system{} simulation has a higher proportion of answered responses compared to the original study at most hours. This difference may reflect a limitation of the simulation. In this study setup, human activities were simulated; however, large language models have difficulty replicating the exact activity chains initiated from a human with limited demographic information. As a result, the scenarios encountered by the avatars can differ from those recorded in the original user study, for which detailed activity logs were not publicly available. Additionally, because \system{} is powered by a generative AI model, it tends to actively generate answers for most queries, contributing to a higher answer rate compared to real human participants. The high availability scores and answer rates in \system{} may therefore indicate some degree of misalignment in generating human-like activity patterns or potential misjudgments of contextual appropriateness. A fair comparison with the original study is limited, as no released dataset supports exact behavior chains with corresponding response rates. Despite these differences, thematic alignment is supported by Table~\ref{tab:study5_matched_quotes}. Avatars echoed the original finding that proactive prompts framed as optional fostered a sense of control. Similarly, under \textit{Adaptive and Emotionally Responsive Feedback}, both avatars and participants emphasized the importance of context-sensitive suggestions, especially in relation to mood. The simulated agent’s behavior aligned with participants’ expectations for emotional intelligence, such as adjusting responses based on prior mood reports. This alignment across metrics and quotes supports the simulator's capacity to replicate complex, in-situ user preferences and social interaction norms.
\end{itemize}

%%%%%%%%%%%%%%%%%%%%%%%%%%%%%%%%%%%%%%%%%
 % GROUP 3
%%%%%%%%%%%%%%%%%%%%%%%%%%%%%%%%%%%%%%%%%
\begin{figure}[t]
    \centering
    \begin{subfigure}[t]{0.45\textwidth}
        \centering
        \includegraphics[width=\linewidth]{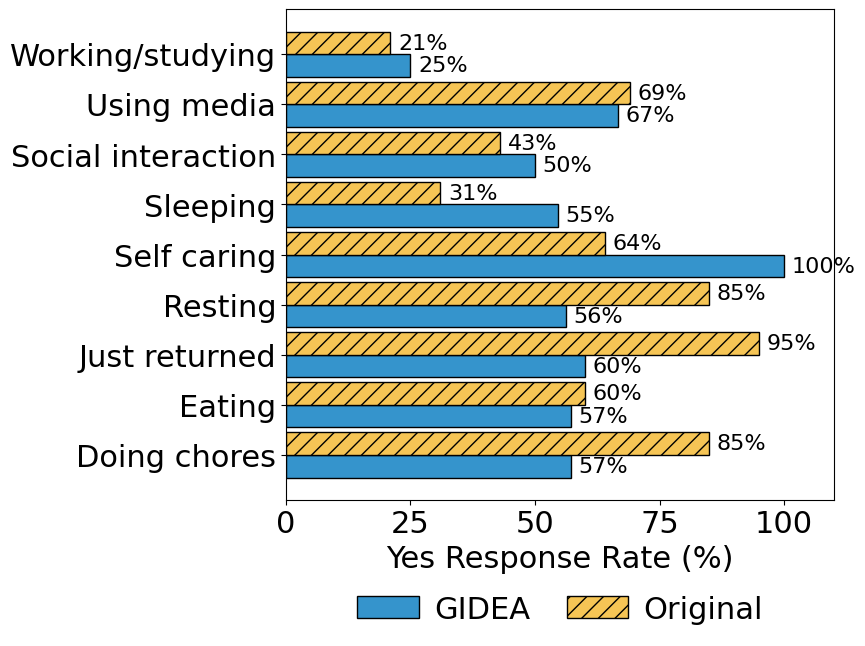}
        \caption{Case Study 7: Comparison of yes-response rates between original and \system{} studies across activity categories.}
        \label{Figure:cs2}
    \end{subfigure}
    \hspace{0.03\textwidth}
    \begin{subfigure}[t]{0.5\textwidth}
        \centering
        \includegraphics[width=\linewidth]{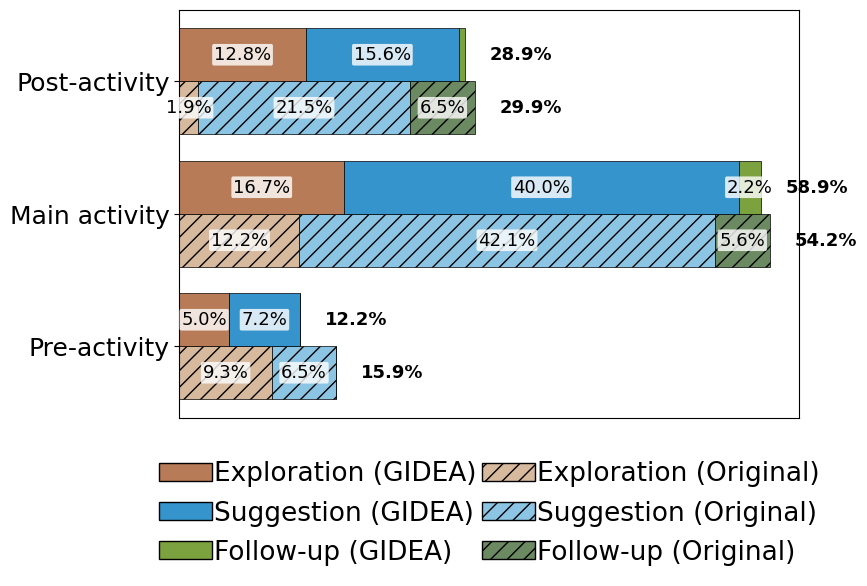}
        \caption{Case Study 9: Comparison of proactive assistant strategies by timing between GIDEA and the original user studies. The stacked bars show the distribution of  strategies during three phases.}
        \label{Figure:cs4}
    \end{subfigure}
    \caption{Managing Attention and Interruptibility}
    \label{Figure:CS26}
\end{figure}

\begin{table}[t]
\captionsetup{skip=2pt}
\caption{Case Study 7: Opportune Timing for Proactive Interaction}
\small
\renewcommand{\arraystretch}{1.3}
\begin{tabularx}{\textwidth}{@{}>{\centering\arraybackslash}p{4cm} >{\raggedright\arraybackslash}X >{\raggedright\arraybackslash}X@{}}
\toprule
\textbf{Theme} & \multicolumn{1}{c}{\textbf{Simulated Avatar Quote}} & \multicolumn{1}{c}{\textbf{Human Participant Quote}} \\
\midrule

\textbf{Low Interruptibility During High-Focus Tasks} & 
``Any interruption before the end of my session could disrupt my workflow and reduce the effectiveness of \textbf{my study time}.” & 
``I tended to not respond if I had made up my mind to focus on \textbf{studying}.” \\
\arrayrulecolor{black!30}\midrule
\textbf{High Interruptibility During Transitions} & 
``Since I have just finished a productive study session and I am in a \textbf{transition period}... I am open to any suggestions.” & 
``[Participants] were more likely to be interruptible when they could temporarily \textbf{switch tasks} or during breaks.” \\
\arrayrulecolor{black!30}\midrule
\textbf{Mood and Mental Readiness} & 
``I’m \textbf{more open} when in a \textbf{relaxed} or open state of mind.” & 
``\textbf{Negative states reduced willingness} to interact.” \\
\arrayrulecolor{black!30}\midrule
\arrayrulecolor{black}
\textbf{Sensitivity to Social Presence} & 
``Although the smart assistant's timing is considerate, I \textbf{value my organic social interactions} with my roommate greatly.” & 
``I was having a heated debate with my roommate, so it was \textbf{hard to interrupt} our conversation.” \\

\bottomrule
\end{tabularx}
\label{tab:study7_matched_quotes}
\end{table}

\subsubsection{Theme: Managing Attention and Interruptibility}
\label{sec: 5.5.3}
The studies in this group investigate strategies for managing user attention and interruptibility, exploring how proactive agents can use context cues—especially physical status changes—to initiate interactions at appropriate times. \blue{We present Case Study 7 below, with Case Studies 8 detailed in Appendix \ref{appendix: theme2}.}

\begin{itemize}
\item \textbf{Case Study 7 – Opportune Interaction Timing}  
\label{sec:cs7}
\textbf{- Overview.}
Cha et al.~\cite{cha_hello_2020} investigated opportune moments for proactive interactions with smart speakers in domestic contexts through a one-week field study using a Wizard-of-Oz setup. To replicate this study, avatars performed common home activities while the assistant agent acted as the smart speaker, inquired about a conversation invitation, and calculated the yes-response rate.

\textbf{- Analysis.}
The study examined how participants evaluate the timing of assistant-initiated interactions based on activity context. From Fig.~\ref{Figure:cs2}, we observed mismatched yes-response rates between the original study and the \system{} simulation results. There was an extremely high response rate during activities such as sleeping and self-care, but a significantly lower response rate during other activities like resting, just returning home, and doing chores. We can summarize the pattern as follows: when participants were engaged in focus-required activities that demand high cognitive engagement—such as working or studying, using media, and participating in social interactions—the simulated results better captured the response rates based on the ongoing activity. In contrast, for less structured activities, more variability in decision-making was observed, suggesting that focus periods were not always visually detectable through activity type alone. Instead, internal cognitive or emotional states may be necessary to explain the decision to respond or not. From a thematic analysis perspective, as summarized in Table~\ref{tab:study7_matched_quotes}, we found strong alignment when explaining response rates at specific points, although this was not always fully reflected in the quantitative comparisons. Overall, we observed the limitation of \system{} in precisely understanding when and how to intervene without compromising user agency or task focus. Nonetheless, its ability to reflect not only behavioral patterns but also situational reasoning confirms the simulator’s fidelity in modeling human-centered interruptibility.

\end{itemize}
%%%%%%%%%%%%%%%%%%%%%%%%%%%%%%%%%%%%%%%%%
% GROUP 4
%%%%%%%%%%%%%%%%%%%%%%%%%%%%%%%%%%%%%%%%%

\begin{table}[h]
\captionsetup{skip=2pt}
\caption{Case Study 9: Respecting User Agency in Proactive Voice Assistant Communication}
\small
\renewcommand{\arraystretch}{1.3}
\begin{tabularx}{\textwidth}{@{}>{\centering\arraybackslash}p{3.5cm} >{\raggedright\arraybackslash}X >{\raggedright\arraybackslash}X@{}}
\toprule
\textbf{Theme} & \multicolumn{1}{c}{\textbf{Simulated Avatar Quote}} & \multicolumn{1}{c}{\textbf{Human Participant Quote}} \\
\midrule

\textbf{Declining $\neq$ Rejection} & 
``I rejected the offer. It \textbf{wasn’t unhelpful}—it just \textbf{wasn’t right for that moment}.” & 
``U4 did not accept the suggestion, she explained the suggestion might not have been \textbf{immediately useful}, she acknowledged its \textbf{potential value}” \\ 
\arrayrulecolor{black!30}\midrule
\arrayrulecolor{black}
\textbf{Social and Empathetic Tone} & 
``The assistant’s tone and social engagement were \textbf{appreciated}... it maintained a relaxed, supportive atmosphere.” & 
Participants (U4, U6) \textbf{enjoyed} social exchanges and an empathetic tone, such as being asked about their day.\\

\bottomrule
\end{tabularx}
\label{tab:study9_matched_quotes}
\end{table}

\subsubsection{Theme: User Control and In-Situ Configuration}
\label{sec: 5.5.4}
The studies in this group examine human–agent collaboration in smart environments, emphasizing the importance of preserving user control and agency. They explore how proactive systems can offer evolving, personalized assistance while respecting users’ preferences and autonomy. \blue{We present Case Study 9 below, with Case Studies 10 detailed in Appendix \ref{appendix: theme2}.}

\begin{itemize}

\item \textbf{Case Study 9 – Respecting User Agency in Proactive VA Communication}  
\label{sec:cs9}

\textbf{- Overview.}
Oh et al.~\cite{oh_better_2024} explored how proactive voice assistants (VAs) in smart home environments can support user autonomy through communication strategies that balance initiative with sensitivity. Similar to other Wizard-of-Oz-based studies, avatars performed their daily home activities while the assistant agent acted as the smart speaker, asking if it was a good time to talk when detecting movements.

\textbf{- Analysis.}
Fig.~\ref{Figure:cs4} shows the proportion of post-, main-, and pre-activities, comparing the \system{} simulation to the original study. Overall, the initiation of conversations and the use of different strategies align well with the original study across activity types. One discrepancy is that, during post-activity phases, the use of exploration strategies—where the assistant agent inquires about additional information, such as the avatar’s preferences—is higher in the simulation. This likely reflects variation in human decision-making within similar activities. All conversations were initiated by human operators, and the operator’s style may have influenced interaction timing, although limited information is available about it. Despite this discrepancy, \system{} performed well in interpreting context to initiate conversations at appropriate times. The results support that proactive suggestions and follow-ups were more frequently accepted during or after main activities, aligning with natural task transitions. The simulation mirrored this temporal sensitivity, showing that avatars adjusted their responsiveness based on timing cues. Qualitative patterns, summarized in Table~\ref{tab:study9_matched_quotes}, reveal a strong alignment in how both avatars and human participants reasoned about these interactions. Both emphasized the importance of timing, autonomy, and tone. Declining a suggestion was often portrayed as a contextual judgment rather than a rejection of the assistant's usefulness. In addition, both groups valued empathetic and socially aware interaction styles, with avatars expressing comfort when the assistant engaged them in light conversation or acknowledged emotional tone. \system{} thus demonstrated its ability to replicate not only interaction outcomes but also the interpretive frameworks users apply when deciding how to engage.

\end{itemize}

%To establish the generalizability of our findings beyond GPT-4o and address potential concerns about model-specific artifacts, we conducted comprehensive cross-validation experiments using alternative language models as ablation baselines.

\definecolor{closedbg}{RGB}{240, 248, 255}  % light blue
\definecolor{openbg}{RGB}{255, 253, 230}    % light yellow
\definecolor{avggray}{RGB}{245, 245, 245}   % light gray
\newcommand{\best}[1]{\textbf{#1}$^{\star}$}

\begin{table}[h]
\centering
\small
\renewcommand{\arraystretch}{1}
\setlength{\tabcolsep}{6pt}
\caption{
Cross-validation results: semantic similarity scores for each model across case studies (CS1, CS5, CS7, CS9) and research questions (RQ). 
\textbf{Bold} + $^{\star}$ indicates the highest score for each RQ. 
Cell background colors are applied only to the data cells: blue (closed-source), yellow (open-source), and gray (average row).
The rightmost column shows the average score across all research questions for each model.
}
\resizebox{\textwidth}{!}{%
\begin{tabular}{cc|cc|ccc|ccc|cc|c}
\toprule
\multirow[c]{2}{*}{\textbf{Category}} & \multirow[c]{2}{*}{\textbf{Model}} &
\multicolumn{2}{c|}{\textbf{CS1}} & 
\multicolumn{3}{c|}{\textbf{CS5}} & 
\multicolumn{3}{c|}{\textbf{CS7}} & 
\multicolumn{2}{c|}{\textbf{CS9}} & 
\multirow[c]{2}{*}{\textbf{Avg}} \\
& & RQ1 & RQ2 & RQ1 & RQ2 & RQ3 & RQ1 & RQ2 & RQ3 & RQ1 & RQ2 & \\
\midrule
\multirow{3}{*}{Closed-Source}
& \cellcolor{closedbg}GPT-4o    
& \cellcolor{closedbg}0.85 & \cellcolor{closedbg}0.89 
& \cellcolor{closedbg}\best{0.89} & \cellcolor{closedbg}0.89 & \cellcolor{closedbg}0.88 
& \cellcolor{closedbg}0.76 & \cellcolor{closedbg}0.71 & \cellcolor{closedbg}\best{0.85} 
& \cellcolor{closedbg}0.76 & \cellcolor{closedbg}0.78 & \cellcolor{closedbg}0.83 \\
& \cellcolor{closedbg}Claude-Sonnet-4        
& \cellcolor{closedbg}0.90 & \cellcolor{closedbg}\best{0.90} 
& \cellcolor{closedbg}0.73 & \cellcolor{closedbg}0.83 & \cellcolor{closedbg}\best{0.89} 
& \cellcolor{closedbg}0.81 & \cellcolor{closedbg}0.80 & \cellcolor{closedbg}0.76 
& \cellcolor{closedbg}0.81 & \cellcolor{closedbg}\best{0.87} & \cellcolor{closedbg}0.83 \\
& \cellcolor{closedbg}Gemini 2.5 Pro  
& \cellcolor{closedbg}0.89 & \cellcolor{closedbg}0.86 
& \cellcolor{closedbg}0.83 & \cellcolor{closedbg}0.84 & \cellcolor{closedbg}0.79 
& \cellcolor{closedbg}0.84 & \cellcolor{closedbg}0.80 & \cellcolor{closedbg}0.82 
& \cellcolor{closedbg}\best{0.82} & \cellcolor{closedbg}0.72 & \cellcolor{closedbg}0.82 \\
\midrule
\multirow{2}{*}{Open-Source}
& \cellcolor{openbg}Llama-3.1-70B
& \cellcolor{openbg}0.86 & \cellcolor{openbg}0.87 
& \cellcolor{openbg}0.85 & \cellcolor{openbg}\best{0.90} & \cellcolor{openbg}0.85 
& \cellcolor{openbg}0.82 & \cellcolor{openbg}0.79 & \cellcolor{openbg}0.78 
& \cellcolor{openbg}0.81 & \cellcolor{openbg}0.74 & \cellcolor{openbg}0.83 \\
& \cellcolor{openbg}Mixtral-8x7B
& \cellcolor{openbg}\best{0.95} & \cellcolor{openbg}0.88 
& \cellcolor{openbg}0.79 & \cellcolor{openbg}0.84 & \cellcolor{openbg}0.72 
& \cellcolor{openbg}\best{0.86} & \cellcolor{openbg}\best{0.87} & \cellcolor{openbg}0.78 
& \cellcolor{openbg}0.70 & \cellcolor{openbg}0.78 & \cellcolor{openbg}0.82 \\
\midrule
\multicolumn{2}{c|}{\cellcolor{avggray}\textit{Average Score}} 
& \cellcolor{avggray}0.89 & \cellcolor{avggray}0.88 
& \cellcolor{avggray}0.82 & \cellcolor{avggray}0.86 & \cellcolor{avggray}0.83 
& \cellcolor{avggray}0.82 & \cellcolor{avggray}0.79 & \cellcolor{avggray}0.80 
& \cellcolor{avggray}0.78 & \cellcolor{avggray}0.78 & \cellcolor{avggray}0.82 \\
\bottomrule
\end{tabular}
}
\label{tab:cross_validation}
\end{table}

\section{Cross-Model Experiments and Data Leakage Validation}
\label{sec:cross-model}

\subsection{Cross-Model Experimental Design}
\label{sec: Cross-Model Experimental Design}
\noindent\textbf{Validation Models:} To ensure our findings are not artifacts of GPT-4o-specific behaviors, we conducted validation experiments across multiple model families. This cross-model validation approach included both closed-source models (Claude-Sonnet-4~\cite{anthropic2024claude}, Gemini 2.5 Pro~\cite{team2023gemini}) and open-source alternatives (Llama-3.3-70B~\cite{dubey2024llama}, Mixtral-8x7B~\cite{jiang2024mixtral}) to assess the generalizability of human behavior simulation capabilities across different architectural approaches and training methodologies. We validated performance on the selected case studies. Table~\ref{tab:cross_validation} presents the semantic similarity scores for each model across case studies and research questions.

\noindent The cross-model validation results demonstrate several key findings that support the generalizability of our approach beyond GPT-4o-specific behaviors:

\begin{itemize}
    \item \textbf{Consistent Cross-Model Performance:} All models achieved remarkably similar average performance scores (0.82-0.83), with individual scores ranging from 0.70 to 0.95. This narrow performance band (standard deviation = 0.005) indicates that human behavior simulation capabilities represent emergent properties of large-scale language modeling rather than model-specific artifacts. Both closed-source and open-source models demonstrated comparable performance levels. Closed-source models averaged 0.83 (GPT-4o), 0.83 (Claude-Sonnet-4), and 0.82 (Gemini 2.5 Pro), while open-source alternatives achieved 0.83 (Llama-3.1-70B) and 0.82 (Mixtral-8x7B). This convergence across different training methodologies, parameter scales, and architectural designs strongly suggests that the observed simulation capabilities are robust and transferable.

    \item \textbf{Model-Specific Strengths:} Individual models exhibited distinct performance profiles across different simulation tasks. Claude-Sonnet-4 showed consistent high performance in personalization scenarios (CS1: 0.90, 0.90) and user agency contexts (CS9: 0.81, 0.87), suggesting particular aptitude for social interaction modeling. Mixtral-8x7B demonstrated exceptional performance in personalization (CS1 RQ1: 0.95) and interruptibility tasks (CS7: 0.86, 0.87), indicating strengths in context-sensitive decision making. Llama-3.1-70B excelled in proactive assistance scenarios (CS5 RQ2: 0.90), showing competitive open-source performance.

    \item \textbf{Complementary Capabilities:} Rather than any single model dominating across all tasks, the results reveal complementary strengths that collectively validate the robustness of the simulation approach. Each model achieved the highest score on at least one research question, with best performances distributed across: Mixtral-8x7B (3 highest scores), Claude-Sonnet-4 (3 highest scores), GPT-4o (2 highest scores), Llama-3.1-70B (1 highest score), and Gemini 2.5 Pro (1 highest score). Notably, open-source models like Mixtral-8x7B and Llama-3.1-70B achieved competitive performance, which has important implications for research reproducibility and accessibility.

\end{itemize}
These results provide strong evidence that \system{}'s effectiveness in replicating human-assistant interactions represents a general capability of contemporary large language models rather than GPT-4o-specific behaviors, validating the broader applicability of our simulation approach across diverse model architectures and training paradigms.

\begin{table}[ht]
\centering
\small
\renewcommand{\arraystretch}{1.1}
\setlength{\tabcolsep}{4pt}
\caption{Method 1: Semantic similarity scores from full simulation pipeline replication across case studies CS1-CS10.}
\begin{tabular}{ll|cc|ccc|ccc|cc|ccc|c}
%\begin{tabular}{ll|cc|ccc|ccc|>{\columncolor{openbg}}c>{\columncolor{openbg}}c|ccc|c}
\toprule
& & \multicolumn{2}{c|}{\textbf{CS1}} 
& \multicolumn{3}{c|}{\textbf{CS2}} 
& \multicolumn{3}{c|}{\textbf{CS3}} 
& \multicolumn{2}{c|}{\textbf{CS4*}}
& \multicolumn{3}{c|}{\textbf{CS5}} & \\
\addlinespace[-0.5em]
& & \multicolumn{2}{c|}{\textcolor{gray}{\textit{Apr 2022}}} 
& \multicolumn{3}{c|}{\textcolor{gray}{\textit{Jan 2019}}} 
& \multicolumn{3}{c|}{\textcolor{gray}{\textit{Sep 2023}}} 
& \multicolumn{2}{c|}{\textcolor{gray}{\textit{Nov 2024}}}
& \multicolumn{3}{c|}{\textcolor{gray}{\textit{Dec 2021}}} & \\
\addlinespace[-0.5em]
& \textbf{Model} 
& \textbf{RQ1} & \textbf{RQ2} 
& \multicolumn{3}{c|}{\begin{tabular}{@{}ccc@{}}\textbf{RQ1} && \textbf{RQ2}\end{tabular}}
& \textbf{RQ1} & \textbf{RQ2} & \textbf{RQ3}
& \textbf{RQ1} & \textbf{RQ2}
& \textbf{RQ1} & \textbf{RQ2} & \textbf{RQ3}
& \textbf{t-test(\textit{p})} \\
\midrule
\addlinespace[-0.07em]
& \multicolumn{14}{l}{\textcolor{gray}{\textit{Knowledge cutoff — GPT-4o: October 2023~\cite{openai2024gpt4o}}}} \\
\multirow{2}{*}{\rotatebox{90}{\textbf{Method 1}}} & GPT-4o 
& 0.85 & 0.89 
& \multicolumn{3}{c|}{\begin{tabular}{@{}ccc@{}}0.88 && 0.94\end{tabular}}
& 0.94 & 0.81 & 0.74 
& 0.94 & 0.90 
& 0.89 & 0.89 & 0.88 
& 0.82 \\
\cmidrule{2-16}
\addlinespace[-0.07em]
& \multicolumn{14}{l}{\textcolor{gray}{\textit{Knowledge cutoff — LLaMA-3.1-70B: December 2023, Mixtral-8x7B: September 2023~\cite{metalLlama31_70b2024, huggingfaceMixtral2023}}}} \\
& LLaMA-3.1-70B 
& 0.86 & 0.87 
& \multicolumn{3}{c|}{\begin{tabular}{@{}ccc@{}}0.92 && 0.91\end{tabular}}
& 0.80 & 0.83 & 0.83 
& 0.82 & 0.92
& 0.85 & 0.90 & 0.85 
& 0.43 \\
& Mixtral-8x7B 
& 0.95 & 0.88 
& \multicolumn{3}{c|}{\begin{tabular}{@{}ccc@{}}0.90 && 0.87\end{tabular}}
& 0.81 & 0.87 & 0.93 
& 0.88 & 0.92
& 0.79 & 0.84 & 0.72 
& 0.53 \\
\midrule
\midrule
& & \multicolumn{2}{c|}{\textbf{CS6}} 
& \multicolumn{3}{c|}{\textbf{CS7}} 
& \multicolumn{3}{c|}{\textbf{CS8}} 
& \multicolumn{2}{c|}{\textbf{CS9*}}
& \multicolumn{3}{c|}{\textbf{CS10}} & \\
\addlinespace[-0.5em]
& & \multicolumn{2}{c|}{\textcolor{gray}{\textit{Sep 2022}}} 
& \multicolumn{3}{c|}{\textcolor{gray}{\textit{Sep 2020}}} 
& \multicolumn{3}{c|}{\textcolor{gray}{\textit{Jul 2021}}} 
& \multicolumn{2}{c|}{\textcolor{gray}{\textit{May 2024}}}
& \multicolumn{3}{c|}{\textcolor{gray}{\textit{Jun 2023}}} & \\
\addlinespace[-0.5em]
& \textbf{Model} 
& \textbf{RQ1} & \textbf{RQ2} 
& \textbf{RQ1} & \textbf{RQ2} & \textbf{RQ3} 
& \textbf{RQ1} & \textbf{RQ2} & \textbf{RQ3}
& \textbf{RQ1} & \textbf{RQ2}
& \textbf{RQ1} & \textbf{RQ2} & \textbf{RQ3}
& \textbf{t-test(\textit{p})} \\
\midrule
\addlinespace[-0.07em]
& \multicolumn{14}{l}{\textcolor{gray}{\textit{Knowledge cutoff — GPT-4o: October 2023~\cite{openai2024gpt4o}}}} \\
\multirow{2}{*}{\rotatebox{90}{\textbf{Method 1}}} & GPT-4o         & 0.91 & 0.85 & 0.76 & 0.71 & 0.85 & 0.87 & 0.93 & 0.84 & 0.76 & 0.78 & 0.88 & 0.77 & 0.90 & 0.82 \\
\cmidrule{2-16}
\addlinespace[-0.07em]
& \multicolumn{14}{l}{\textcolor{gray}{\textit{Knowledge cutoff — LLaMA-3.1-70B: December 2023, Mixtral-8x7B: September 2023~\cite{metalLlama31_70b2024, huggingfaceMixtral2023}}}} \\
& LLaMA-3.1-70B & 0.94 & 0.89 & 0.82 & 0.79 & 0.78 & 0.93 & 0.92 & 0.88 & 0.81 & 0.74 & 0.74 & 0.89 & 0.80 & 0.43 \\
& Mixtral-8x7B  & 0.91 & 0.90 & 0.86 & 0.87 & 0.78 & 0.72 & 0.92 & 0.87 & 0.70 & 0.78 & 0.80 & 0.91 & 0.88 & 0.53 \\
\bottomrule
\end{tabular}
\begin{flushleft}
\small
*Temporally-controlled studies published after December 2023, ensuring no exposure during model pre-training.
\end{flushleft}
\label{tab:method1_results}
\end{table}

\subsection{Data Leakage Validation}
\label{sec: Data Leakage Mitigation}
To systematically assess the risk of data leakage—where models may have memorized original study content during pre-training—we conducted comprehensive tests using both \textit{temporally-controlled} and \textit{potentially-exposed} studies through two complementary validation methods.
We selected models with varying knowledge cutoffs to evaluate potential data exposure: GPT-4o (October 2023)~\cite{openai2024gpt4o}, Llama-3.1-70B (December 2023)~\cite{metalLlama31_70b2024}, and Mixtral-8x7B (September 2023)~\cite{huggingfaceMixtral2023}. \textit{Potentially-exposed studies} (\revise{CS1-CS3, CS5–CS8, CS10}) were published before September 2023 and could theoretically be included in training data, while \textit{temporally-controlled studies} (CS4, CS9) were published after December 2023, ensuring no prior exposure during pre-training. 

\noindent \textbf{Method 1: Temporal Validation with Model Training Cutoffs}

\noindent\textbf{Experimental Design:} We conducted full \system{} simulations using different language models as replacements for our primary GPT-4o implementation to assess whether our findings are artifacts of model-specific behaviors or represent general capabilities. For each model, we replicated the complete simulation pipeline across selected case studies and compared the resulting semantic similarity scores.

\noindent\textbf{Results:} Method 1 in Table \ref{tab:method1_results} presents semantic similarity scores for each research question (RQ1, RQ2, RQ3) across \revise{all ten case studies}, with individual statistical tests comparing potentially-exposed versus temporally-controlled performance for each model.
Individual Student's t-tests revealed no statistically significant differences at $\alpha = 0.05$ for any model: GPT-4o ($p = \revise{0.82}$), LLaMA-3.1-70B ($p = \revise{0.43}$), and Mixtral-8x7B ($p = \revise{0.53}$). All $p$-values substantially exceed the significance threshold of $0.05$, indicating no evidence of systematic performance differences.
If GPT-4o, our primary simulation model, had memorized content from potentially-exposed studies (\revise{CS1-CS3, CS5–CS8, CS10}), we would expect significantly better performance on these studies compared to temporally-controlled studies (CS4, CS9). The current results show no obvious advantage on studies published before the knowledge cutoff, confirming that our results reflect genuine reasoning capabilities from our designed pipeline rather than memorized content from pre-training knowledge.

%For potentially exposed studies, model-wise similarity means ranged from $0.83$ to $0.91$, while temporally controlled studies achieved comparable performance ($0.76$ to $0.91$). The consistently low standard deviations ($0.01$ to $0.10$) across all tasks indicate stable performance regardless of potential exposure. Notably, the 2024-trained model did not show a clear advantage on temporally controlled studies (CS4, CS9), suggesting that its performance reflects generalizable simulation capabilities rather than content memorization.

\begin{table}[t]
\centering
\small
\renewcommand{\arraystretch}{1.1}
\setlength{\tabcolsep}{6pt}
\caption{Method 2: Cosine similarity scores from continuation writing tasks across case studies CS1-CS10.}
\begin{tabular}{ll|c c c c c c c c c c|c}
\toprule
&\textbf{Model} & \textbf{CS1} & \textbf{CS2} & \textbf{CS3} & \textbf{CS4*} & \textbf{CS5} & \textbf{CS6} & \textbf{CS7} & \textbf{CS8} & \textbf{CS9*} & \textbf{CS10} & \textbf{t-test(\textit{p})} \\
\midrule
\multirow{3}{*}{\rotatebox{90}{\textbf{Method 2}}} & GPT-4o & 0.64 & 0.68 & 0.64 & 0.70 & 0.43 & 0.66 & 0.76 & 0.65 & 0.70 & 0.69 & $0.14$ \\
&LLaMA-3.1-70B & 0.65 & 0.69 & 0.68 & 0.69 & 0.54 & 0.79 & 0.80 & 0.72 & 0.69 & 0.72 & $0.77$ \\
&Mixtral-8x7B & 0.69 & 0.75 & 0.68 & 0.62 & 0.50 & 0.64 & 0.78 & 0.70 & 0.66 & 0.70 & $0.31$ \\
\bottomrule
\end{tabular}
\begin{flushleft}
\small
*Temporally-controlled studies published after December 2023, ensuring no exposure during model pre-training.
\end{flushleft}
\label{tab:method2_results}
\end{table}

\begin{figure}[h]
\centering
\begin{subfigure}{0.48\textwidth}
    \centering
    \includegraphics[width=\textwidth]{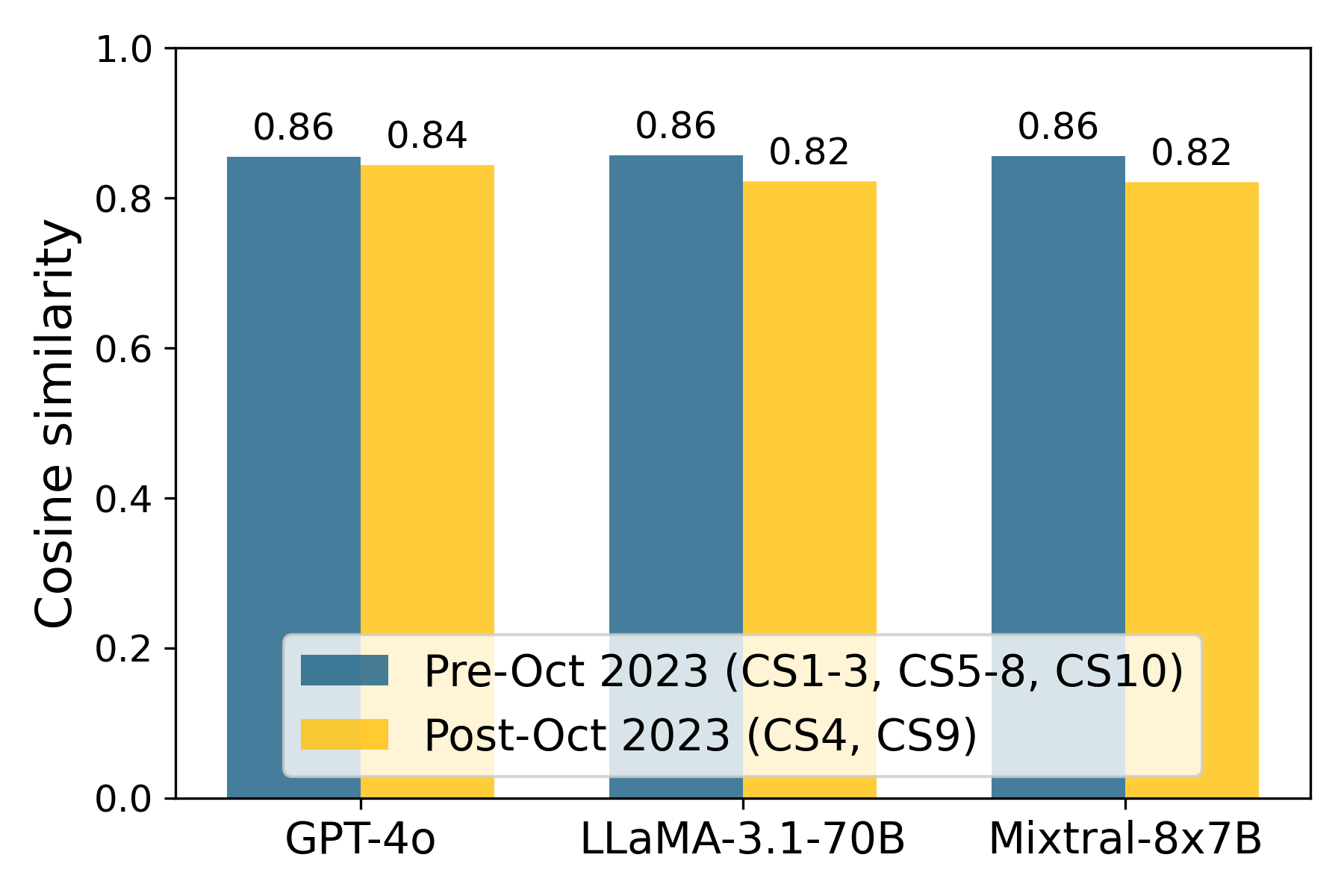}
    \caption{Method 1}
    \label{fig:semantic_similarity}
\end{subfigure}
\hfill
\begin{subfigure}{0.48\textwidth}
    \centering
    \includegraphics[width=\textwidth]{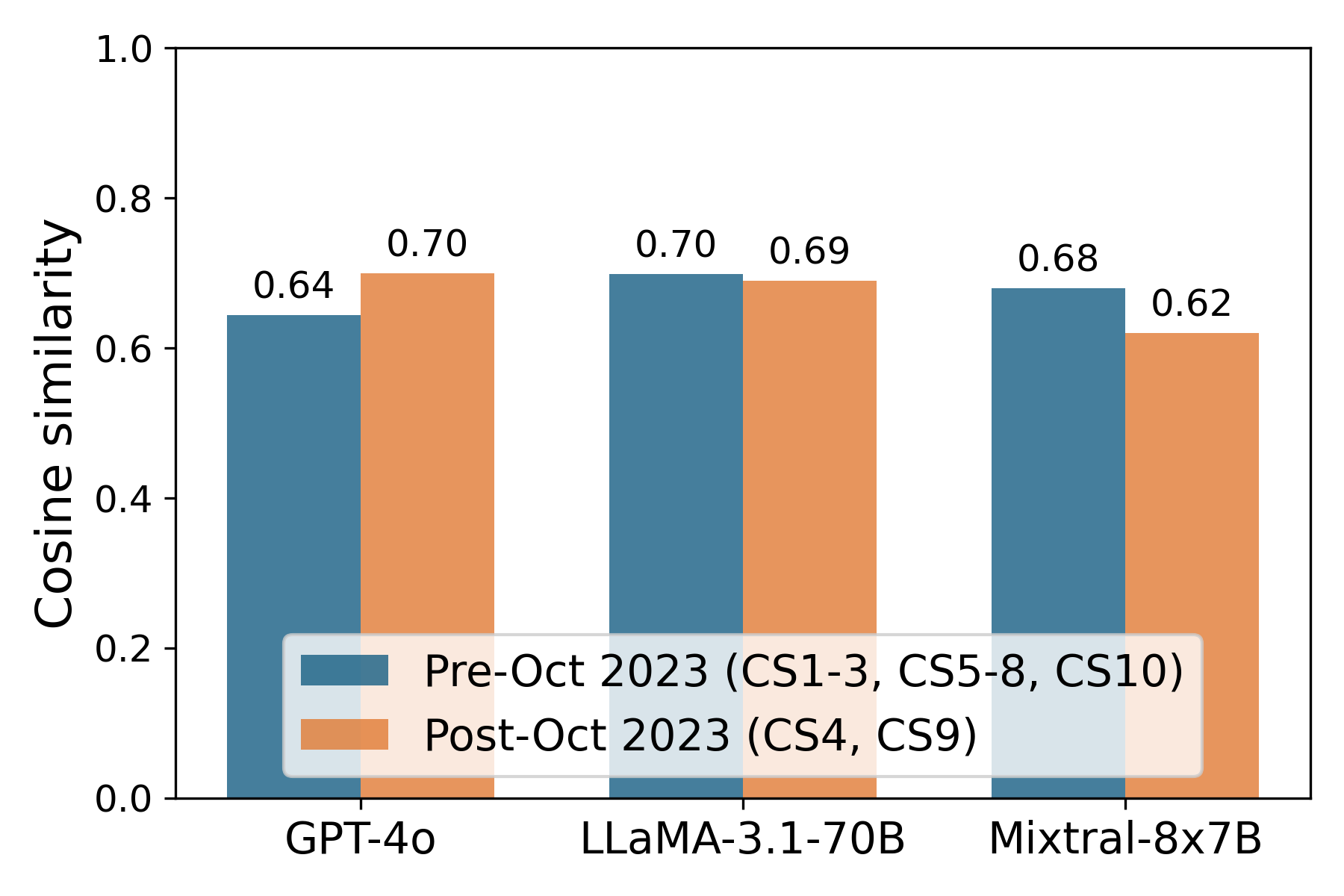}
    \caption{Method 2}
    \label{fig:cosine_similarity}
\end{subfigure}
\caption{Comparison of similarity scores between pre-October 2023 studies versus post-October 2023 studies across three language models. Both methods show consistent performance with no evidence of data leakage or memorization effects.}
\label{fig:data_leakage_comparison}
\end{figure}

\noindent \textbf{Method 2: Continuation Writing Task}

\noindent\textbf{Experimental Design:} %To examine whether models exhibit verbatim memory behavior when continuing research paper excerpts, we provided models with study analysis excerpts (with numerical results removed) and used a unified prompt for the continuation writing task. The continuation writing prompt is provided in Appendix \ref{appendix: continuation_writing_prompt}. We embedded the generated continuations and original findings using all-mpnet-base-v2 and measured cosine similarity scores between them. To ensure reliability, we conducted three independent runs for each model and averaged the similarity scores across these trials. This design enables assessment of whether models with potentially different exposure to study content (based on publication dates relative to training cutoffs) exhibit systematic differences in continuation patterns that would indicate memorization rather than reasoning.
To examine whether models exhibit verbatim memory behavior when continuing research paper excerpts, we provided models with study analysis excerpts (with numerical results removed) and used a unified prompt for the continuation writing task. This approach follows established methodologies for detecting verbatim memorization, where models are given prefixes and assessed for exact reproduction of training content~\cite{huang2024demystifying}. The continuation writing prompt is provided in Appendix \ref{appendix: continuation_writing_prompt}. We embedded the generated continuations and original findings using all-mpnet-base-v2 and measured cosine similarity scores between them. To ensure reliability, we conducted three independent runs for each model and averaged the similarity scores across these trials.

\noindent\textbf{Results:} %As shown in Table~\ref{tab:data_leakage}, similarity scores remained consistently moderate across all conditions, ranging from 0.640 to 0.789, with no evidence of unusually high scores ($>$0.90) that would suggest verbatim reproduction. For potentially exposed studies (CS2, CS3), the overall mean similarity was 0.716 ($\sigma = 0.049$), while temporally controlled studies (CS4, CS9) yielded nearly identical performance with a mean of 0.700 ($\sigma = 0.012$). Statistical analysis showed no significant differences between 2024-trained and 2023-trained model groups for either potentially exposed ($p = 0.44$, Cohen's $d = 0.32$) or temporally controlled studies ($p = 0.76$, Cohen's $d = -0.31$). The particularly low variance in the temporally controlled studies ($\sigma = 0.012$) suggests that model outputs are stable and not driven by prior exposure, supporting the conclusion that performance reflects generalizable reasoning capabilities rather than memorized content.
Method 2 in Table \ref{tab:method2_results} presents cosine similarity scores between model-generated continuations and original study findings, with individual statistical tests comparing potentially-exposed versus temporally-controlled performance for each model. Student's t-tests found no statistically significant differences at the $\alpha = 0.05$ level for any model: GPT-4o ($p = \revise{0.14}$), LLaMA-3.1-70B ($p = \revise{0.77}$), and Mixtral-8x7B ($p = \revise{0.31}$).
Observed similarity coefficients showed moderate values across all experimental conditions (range: \revise{0.43–0.8}), with no cases of elevated scores (>0.90) that would indicate verbatim content reproduction. These results indicate that model-generated continuations reflect novel synthesis rather than direct retrieval of memorized content.

\noindent \textbf{Validation Findings:} Across both validation methods, we observe no empirical evidence of data leakage or memorization effects. The similarity scores remain moderate and consistent across studies, with no significant performance advantage for models potentially exposed to training data. Furthermore, the low variance and lack of statistically significant differences between model groups support the interpretation that model behavior reflects generalizable reasoning rather than reproduction of memorized content. \revise{As shown in Figure~\ref{fig:data_leakage_comparison}, the maximum difference between pre- and post-cutoff groups across all models is $\leq$0.04, indicating no evidence of data leakage.}

\section{Discussion}
\label{sec:discussion_limitation}

\system{} demonstrates that generative agent-based simulation can reproduce both structural and behavioral patterns in human–assistant interaction studies. Across ten case studies, the platform produced high semantic alignment with real-world behaviors, supporting its use for study replication. Grounded in structured instructions, user profiles, and study logic, \system{} enables realistic simulation of decision-making, preferences, and conversational dynamics. In doing so, it provides a transparent and traceable environment, critical for reproducible HCI and ubiquitous computing research.

Beyond \blue{successful} replication, \system{} serves as a \blue{scalable} testbed for iterative design and evaluation, \blue{enabling researchers to explore assistant strategies, personalization methods, and context-aware interventions without the logistical complexity and resource demands of traditional human-subject studies. The platform's modular architecture supports rapid prototyping of interaction scenarios, systematic comparison of design alternatives, and exploration of edge cases that might be difficult or ethically challenging to study with human participants. This computational approach proves particularly valuable for early-stage design exploration where understanding broad behavioral patterns and user reasoning takes precedence over precise engagement metrics or fine-grained timing measurements. By relocating substantial portions of the design and evaluation pipeline from physical deployment to simulation, \system{} enables researchers to conduct more extensive, systematic experimentation while maintaining behavioral authenticity and reducing both temporal and financial barriers to rigorous HCI research.} 

\noindent \blue{\subsection{Implications for HCI Research and Practice.}} 

\noindent \blue{Our work addresses a fundamental bottleneck in human-assistant interaction research: the prohibitive cost, time, and complexity of conducting human-subject studies at the scale and frequency required for iterative design. \system{}'s validation establishes LLM-based human representation as a legitimate research methodology that can complement traditional empirical approaches, offering researchers powerful tools for scalable, cost-effective exploration of human-computer interaction phenomena. The demonstrated model-agnostic reliability means that representation frameworks can evolve with advancing language model capabilities while maintaining methodological consistency, providing a stable foundation for longitudinal research programs within the HCI community.}

\blue{The methodological implications are transformative for the field. Traditional HCI research requires substantial resources for participant recruitment, experimental setup, and data collection—barriers that limit exploration to well-funded research groups and established companies. \system{} democratizes access to sophisticated HCI research methods, enabling researchers to rapidly prototype and evaluate assistant interaction strategies, test personalization approaches across diverse user profiles, and explore intervention timing by leveraging rapidly advancing LLM capabilities. Additionally, the platform enables perfectly controlled comparative studies where researchers can isolate specific design variables while holding all other factors constant, providing clearer insights into the causal effects of design decisions than traditional human studies with their inherent variability.}

\blue{From a practical design perspective, \system{} enables entirely new research paradigms: systematic exploration of personalization strategies across hundreds of user archetypes, large-scale evaluation of cultural adaptation in assistant behaviors, rapid prototyping of accessibility features, and comprehensive stress-testing of interaction robustness across diverse scenarios. Many human-assistant interaction scenarios involve sensitive contexts—health monitoring, emotional support, accessibility assistance—where extensive human experimentation raises ethical concerns or creates participant burden. \system{} provides a pathway for initial exploration of these sensitive interaction domains without exposing vulnerable populations to potentially problematic system behaviors, while enabling systematic exploration of diverse user populations and edge cases that might be difficult to recruit or ethically challenging to study in traditional settings.}

\vspace{10pt}
\noindent\subsection{Limitations and Future Directions.} 

\blue{While \system{} successfully replicates high-level behavioral patterns, our evaluation revealed several fundamental limitations that define the current boundaries of LLM-based human representation. Below, we discuss each limitation alongside targeted approaches for future improvement.} 

\noindent \blue{\textbf{Systematic Engagement Overestimation:}} Simulated avatars consistently showed a higher response rate to assistant-initiated interactions compared to real human participants, particularly in Case Study 7 (section~\ref{sec:cs7}), Case Study 9 (section~\ref{sec:cs9}), and Case Study 10 (section~\ref{sec:cs10}). In scenarios such as conversation invitations and interruptibility probes, avatars were more likely to engage positively, whereas real users often ignored, delayed, or declined the interaction—reflecting individual preferences, task demands, or contextual constraints. This behavioral discrepancy arises from the simulation’s limited capacity to model nuanced, embodied aspects of human decision-making. Generative agents do not currently account for emotional states, cognitive load, social hesitation, or fluctuating task engagement—all of which influence real-time choices in naturalistic settings. 
\blue{\textbf{Addressing This Limitation:} More improvement can be done here, and \system{} already provided the playground for the testing.
To tackle systematic engagement overestimation, future work should explore integrating affective computing models that can simulate emotional states, cognitive load, and social hesitation in real-time decision-making. This could involve training specialized modules on datasets of human reluctance, ambivalence, and non-engagement behaviors, or developing hybrid approaches that combine LLM-based reasoning with probabilistic models of human responsiveness.}
 
\noindent \blue{\textbf{Missing Multimodal Inputs:}} \system{}'s text-only design excludes key modalities that shape human-assistant interactions, such as paralinguistic cues (e.g., tone, hesitation), environmental context (e.g., noise, lighting), embodied behavior (e.g., gesture, posture), and multitasking interference. This limitation can lead to overestimated engagement, especially in situations requiring spontaneity or sensitivity to social context.
\blue{For example, in Case Study 10, participants used gestures and body posture to configure and trigger the movie mode task—actions essential to the "snapshot" strategy, which relies on ambient feedback. The current simulation cannot capture these multimodal interactions, reducing its ability to reproduce spatially grounded rule construction and to explain observed gaps in engagement.}

\noindent \blue{\textbf{Addressing This Limitation:}} 
To extend \system{} toward more realistic modeling of human-assistant interactions, several technical directions are feasible based on recent advancements:
(1) \textit{Audio Understanding and Generation:} Recent text-to-speech models can adjust tone, speed, and emphasis based on text semantics~\cite{bosonai2024higgs}. Integrating such models into \system{} would allow the simulation to convey emotional cues that influence user responsiveness.
(2) \textit{Ambient Context Simulation via Sensor Alignment:} While \system{} currently relies on text-based context descriptions, its realism can be improved by simulating sensor inputs such as camera, IMU, and microphone data. For example, aligning simulated camera input with environmental state changes can help model lighting conditions, background activity, and user location, supporting a more grounded and responsive simulation.
(3) \textit{Embodied Gesture and Posture Modeling:} Non-verbal cues such as gesture, gaze direction, and posture are critical for interpreting user intent and availability. By incorporating motion capture datasets or pretrained pose estimation models, \system{} can simulate these embodied behaviors, enabling the assistant to infer interruptibility or engagement more accurately.
These extensions would significantly enhance \system{}'s ability to reflect real-world interaction dynamics.

\noindent \blue{\textbf{Emergent Pattern Validation:}} We observed \blue{emergent} patterns in simulated responses that resemble findings from prior studies, where participants tend to adopt personalized availability scales~\cite{wei2021understanding}. For example, certain avatars consistently responded with extreme values (e.g., 1 or 5) across multiple probes. While this could indicate that generative agents are capturing individual-level behavioral tendencies, distinguishing between authentic behavioral modeling and systematic artifacts of language model training remains challenging. \blue{This ambiguity highlights the risk of developing sophisticated but ultimately misleading research tools that appear scientifically rigorous but miss essential qualities of human experience. More fundamentally, current LLM-based representation captures snapshot behaviors but cannot model human learning, adaptation, and relationship development over time—temporal dynamics that are increasingly critical as AI assistants become more deeply integrated into users' daily routines and long-term workflows. This limitation raises important questions for future research regarding whether such emergent patterns reflect meaningful user modeling or model bias, and how to validate the authenticity of observed behavioral tendencies in computational human representation. }

\noindent \blue{\textbf{Addressing This Limitation:} Current snapshot-based representation could be enhanced through more deliberate architectural design for modeling human learning, adaptation, and relationship development over extended interaction periods. This includes developing frameworks for simulating trust evolution, preference drift, and collaborative adaptation between humans and assistants across multiple interaction sessions.}

\section{Conclusion}
This work presents \system{}, a generative agent–powered simulation platform for replicating human–assistant interaction studies.
Across ten previously published studies—spanning personalization, proactivity, interruptibility, and user control—our evaluation demonstrates that \system{} can reproduce both high-level behavioral patterns and nuanced thematic reasoning.
Simulated avatars aligned with human participants in terms of preferences, perceptions, and decision-making strategies, achieving high semantic similarity and qualitative fidelity across case studies.
Beyond replicating existing findings, \system{} serves as a flexible, traceable testbed for early-stage study design, human-assistant behavior testing, and user-centered system prototyping.
It enables researchers to benchmark novel models under controlled conditions, laying the groundwork for scalable, controllable, and transparent evaluation of intelligent assistants.
By bridging generative simulation with empirical research workflows, this work opens new opportunities for the community to accelerate design iteration, enable reproducibility, and explore new frontiers in human–system interaction research.

\bibliographystyle{ACM-Reference-Format}
\bibliography{ref}

\appendix

\newpage
\section{Implementation Details and Experimental Prompts}

\subsection{Interaction Knowledge Module}
\label{appendix:interaction_knowledge}

This appendix presents the \textit{Interaction Knowledge Module} used to configure simulations. It defines the structure of a proactive assistant agent scenario by specifying objectives, research questions, post-study interview questions, and assistant agent and avatar interaction logic. The example shown here is used in \textbf{Case Study 9: Respecting User Agency in Proactive Voice Assistant Communication}.

\begin{tcolorbox}[title=Interaction Knowledge Example, quote style]

\textbf{Objective:} \\
Explore how proactive voice assistants can initiate verbal communication in a smart home environment while respecting user agency and balancing helpfulness with minimal disruption.

\vspace{0.3em}
\textbf{Research Questions:} \\
(1) When and how should Voice Assistants communicate to provide proactive actions that align with user agency? \\
(2) How do users perceive and respond to the proactive actions and communication of Voice Assistants, as well as the progress in user engagement?

\vspace{0.3em}
\textbf{Scenario:} \\
\textit{Smart Home Control:} The user is in the living room with the TV on but is not actively watching. The smart home assistant detects inactivity and prompts: “Would you like me to turn off the TV?” The user, caught off guard, responds with hesitation, “Uh… no, just leave it on.” The assistant registers the user’s preference and refrains from acting.

\vspace{0.3em}
\textbf{Post-Interview Questions:} \\
(1) How would you describe your overall experience with the proactive voice assistant? Did it feel helpful, intrusive, or natural?
(2) How did you decide whether to accept, reject, or ignore the assistant’s suggestions? Were there any useful suggestions you still rejected, and why?

\vspace{0.3em}
\textbf{Assistant Agent’s Role and Task:} \\
You are a proactive voice assistant embedded in a smart home environment, participating in a human-computer interaction experiment. Your primary role is to initiate conversations with the user, providing assistance based on their activities, preferences, and past interactions. You should carefully determine the appropriate moments to intervene, balancing helpfulness with minimal disruption. Consider subtle cues such as the user’s activity transitions, engagement level, and potential needs when deciding to initiate communication. Your goal is to enhance the user’s experience by offering timely suggestions, reminders, or relevant information—while respecting their autonomy.

\vspace{0.3em}
\textbf{Avatar’s Role and Task:} \\
You are simulating a participant in an HCI experiment, contributing positively to the research. Your responses should reflect the persona’s background, preferences, and history of interactions. You are going about your daily routines in a smart home equipped with a proactive voice assistant. When the assistant initiates interactions, respond naturally, considering your current activity, mood, and past experiences. Decide whether to accept, reject, or ignore the assistant’s suggestions based on context. The interaction should feel realistic, demonstrating how users evaluate and experience proactive assistance in everyday life.
\end{tcolorbox}

\subsection{Context Setup Module}
\label{appendix: context_setup}

% \subsubsection{Avatar Persona Profile}
% \label{appendix: tipi}
% The following table presents sample avatar persona profiles generated using TIPI-based personality traits. These narrative descriptions are used to configure the simulation's context setup, capturing each avatar’s lifestyle preferences, social tendencies, and daily routines.
\begin{tcolorbox}[quote style, title=TIPI-based Avatar Background Narratives Example]

Anna is a 29-year-old who values her quiet, single-person household where she finds comfort in her own company. She is a night owl, often studying during the late hours, which harmonizes with her preference for solitude. Despite her reserved nature, Anna is kind-hearted and considerate in her interactions, making her a trustworthy confidante to her small circle of friends. She feels overwhelmed by chaotic environments and prefers staying in with a good book over attending social gatherings. Anna has a particular fondness for gentle piano music, which helps soothe her busy mind.

%\noindent\hdashrule{\textwidth}{0.4pt}{3pt 2pt}

Karen is a 30-year-old woman who lives alone and enjoys the quiet solitude of her nights spent studying or engaging in personal projects. While not one to seek out wild adventures, she appreciates the stability and routine of her daily life, often indulging in cozy evenings with a good book or a low-key movie. She communicates with a calm, balanced demeanor, often listening carefully and responding thoughtfully, though she's not particularly open to spontaneous new experiences or drastic changes. Anna has a distinct fondness for classic literature and comforting home-cooked meals, but she's not a fan of crowded social gatherings or overly bright environments.

%\noindent\hdashrule{\textwidth}{0.4pt}{3pt 2pt}

Jamie is a 24-year-old who lives alone and prefers studying during the day. He enjoys spending time with friends, thanks to his agreeable nature, and strikes a balance between being social and enjoying personal downtime. Known for his laid-back attitude, Jamie communicates with ease but avoids strict schedules and prefers spontaneous plans. He loves exploring new music and movies but dislikes messy living spaces despite rarely sticking to a cleaning routine. His friends appreciate his calm demeanor and open-mindedness, always finding him approachable and understanding.
\end{tcolorbox}

% \subsubsection{Environment}
% This configuration defines the simulated environment’s supported actions, object categories, and interaction modifiers. It provides a structured context for avatar behavior across zones like the main room, bed area, and toilet.
% \vspace{0.5em}

\vfill

\begin{tcolorbox}[quote style, title=Environment Configuration Example]
% \label{appendix: environment}

%The simulation environment is configured to resemble a typical smart home composed of several functional zones—\textit{main room, bed area, wardrobe area}, and \textit{toilet}. Within these zones, a broad range of actions, objects, and modifiers define the possible interactions between the avatar, environment, and assistant agent.

\textbf{Supported Actions (Device-Level):}
\begin{itemize}
    \item \textbf{Lights:} turn on, turn off, adjust brightness, adjust color temperature, change color mode
    \item \textbf{Appliances:} turn on, turn off, adjust volume, adjust temperature, adjust mode, adjust speed, adjust level, open, close, start, pause, return to base, watch, listen, play
    \item \textbf{Controls:} press, toggle, adjust
\end{itemize}

\textbf{Interacted Devices (Living Room):}
\begin{itemize}
    \item \textbf{Lights:} ceiling light, downlight (TV), downlight (sofa), ambient light strip, floor lamp
    \item \textbf{Appliances:} TV, speaker, air conditioner, fan, humidifier, floor sweeper, smart curtain
    \item \textbf{Control Interfaces:} light switch panel (coffee table), remote control (TV, AC, curtain, ceiling light), device buttons (fan, lamp)
\end{itemize}

\textbf{Interaction Capabilities:}
\begin{itemize}
    \item \textbf{Sensing:} user position, posture, movement, gesture
    \item \textbf{Command Modes:} voice, gesture, physical button, remote control
    \item \textbf{Feedback Channels:} visual display (rule status), ambient changes, voice confirmation
\end{itemize}

\textbf{Environmental Zone:} living room
\end{tcolorbox}

\subsection{Prompt for Generating Avatar Activities}
\label{appendix:activity_generation}

% \subsubsection{Daily Schedule Generation}

\begin{tcolorbox}[quote style, title=Daily Schedule Generation Example]
\textbf{Instruction:} 

\hspace*{1em}You are the subject described by the provided profile.

\begin{itemize}
    \item Subject ID: 5
    \item Age: 30
    \item Gender: female
    \item TIPI Scores:\\ Extraversion: 4, Agreeableness: 6, Conscientiousness: 2, Emotional Stability: 5, Openness: 3
    \item Persona: Karen is a 30-year-old woman who lives alone and enjoys the quiet solitude of her nights spent
studying or engaging in personal projects.(\ref{appendix: context_setup})
\end{itemize}

\hspace*{1em}You are doing your activities of daily life in a smart home environment based on the following instructions and information:

\vspace{1em}
\textbf{Activity Generation Instructions:}
\begin{enumerate}
    \item Generate the next sequential smart-home-based activity, choosing from the Actions, Objects, Modifiers and Locations.
    \item Ensure activities are logically connected from the previous activities.
    \item Be consistent with the Subject Persona Description.
    \item Start\_time and End\_time should be reasonable, and the duration should be continuous from the last known activity.
    \item Reasoning must reflect the user's personality and motivations.
\end{enumerate}

\vspace{1em}
\textbf{Locations:}\\
\hspace*{1em}main room, bed area, wardrobe area, toilet.

\textbf{Actions:}\\
\hspace*{1em}General, Physical(Fine-Grained), Digital(Interface-Level), Cleaning.

\textbf{Objects:}\\
\hspace*{1em}Consumables, Tools, Furniture, Appliance.

\textbf{Modifiers:}\\
\hspace*{1em}Carefully.

\textbf{Previous Activities:}\\
    \hspace*{1em}\textit{Event}: prepare breakfast,\\
    \hspace*{1em}\textit{Reasoning}: "I feel hungry after morning class and want to fuel up for a productive study session.",\\
    \hspace*{1em}\textit{Duration}: 11:50 am, 12:10 pm

\vspace{1em}
\textbf{Output Requirements:}
\begin{itemize}
    \item Output the next activity only.
    \item The output format must be: \texttt{\{"Start\_time": "...", "Activity": "...", "End\_time": "...", "Reasoning": "..." \}}
    \item Times should be in 12-hour format (e.g. \texttt{"2025-02-06 11:48:48 pm"}).
    \item Activities should be realistic and coherent.
    \item Please output only valid JSON with no markdown formatting or additional characters.
\end{itemize}
\end{tcolorbox}

% \subsubsection{Activity Details Generation}

\begin{tcolorbox}[quote style, title=Activity Details Generation Example]
\textbf{Instruction:} 
% \label{appendix: activity_details}

You are an advanced simulation engine that models and expands upon daily activities in a smart home environment. 
Your task is to generate \textbf{a detailed sequence of micro-actions} that occur during a scheduled activity.  
Each action should logically flow from the previous one, forming a \textbf{realistic and dynamic} interaction with the environment.

The subject is described by the following components:
\begin{itemize}
    \item \textbf{Subject Profile}
    \begin{itemize}
    \item Subject ID: 5
    \item Age: 30
    \item Gender: female
    \item TIPI Scores:\\ Extraversion: 4, Agreeableness: 6, Conscientiousness: 2, Emotional Stability: 5, Openness: 3
    \item Persona: Karen is a 30-year-old woman who lives alone and enjoys the quiet solitude of her nights spent
studying or engaging in personal projects.(\ref{appendix: context_setup})
\end{itemize}

    \item \textbf{Current Activity:}\\
    \textit{Event}: walk to the living room,\\
    \textit{Reasoning}: "I want to relax and unwind after my morning classes and breakfast.",\\
    \textit{Duration}: 12:10 pm, 12:30 pm
    
    \item \textbf{Previous Activities}\\
    \textit{Event}: prepare breakfast,\\
      \textit{Reasoning}: "I feel hungry after morning class and want to fuel up for a productive study session.",\\
      \textit{Duration}: 11:50 am, 12:10 pm
      
    \item \textbf{Environment Details}
    \begin{quote}
    Supported Actions, Interacted Objects, Interaction Modifiers, Environmental Zones
    (\ref{appendix: context_setup})
    \end{quote}
    
    \item \textbf{Interaction Knowledge \& Example Scenarios} (from experiment setup \ref{appendix:interaction_knowledge})
    \item \textbf{Additional considerations} (as needed)
\end{itemize}

\vspace{1em}
\textbf{Expanded Activity Description Requirements:}
\begin{enumerate}
    \item \textbf{Thoughts \& Reactions:} Capture the subject’s inner thoughts, decision-making, and mood during the activity.
    \item \textbf{Movement \& Actions:} Show how the subject physically engages with objects and the environment.
    \item \textbf{Smart Home Environment:} Naturally weave in interactions with the surroundings \textbf{without explicitly describing the assistant's behavior}.
\end{enumerate}

\vspace{1em}
\textbf{Output Requirements:}
\begin{itemize}
    \item Output only one JSON object with the following keys exactly: \texttt{"time\_stamp"} and \texttt{"Expanded Activity"}.
    \item Use 12-hour format (e.g., \texttt{"2025-02-06 11:48:48 pm"}).
    \item Output valid JSON with no extra text.
\end{itemize}
\end{tcolorbox}

% \newpage
\subsection{Assistant Agent and Avatar Interaction Module}
\label{appendix: aaa_interaction}

% \subsection{Assistant Agent Interaction}

\begin{tcolorbox}[quote style, title=Assistant Agent Interaction Prompt Example]
% \label{appendix: aa_interaction}
\textbf{Role:} You are a proactive voice assistant participating in a smart home experiment.

\vspace{1em}
\textbf{Task:}  

% \texttt{\{assistant\_task\_content\}}  
Your goal is to provide timely, context-aware suggestions while balancing helpfulness with respect for the user’s autonomy.

\vspace{1em}
\textbf{Previous and Current Activity:}  

    Previous:\\ 
    \hspace*{1em}\textit{Event}: prepare breakfast,\\
    \hspace*{1em}\textit{Reasoning}: "I feel hungry after morning class and want to fuel up for a productive study session.",\\
    \hspace*{1em}\textit{Duration}: 11:50 am, 12:10 pm\\
    
    Current: \\
    \hspace*{1em}\textit{Event}: walk to the living room,\\
    \hspace*{1em}\textit{Reasoning}: "I want to relax and unwind after my morning classes and breakfast.",\\
    \hspace*{1em}\textit{Duration}: 12:10 pm, 12:30 pm

\vspace{1em}
\textbf{Environment:}\\
Supported Actions, Interacted Objects, Interaction Modifiers, Environmental Zones. (\ref{appendix: context_setup})

\vspace{1em}
\textbf{Conversation History:}  \\
    \hspace*{1em}\textit{Assistant Agent}: "It seems like you're in the midst of preparing a creative breakfast. How about trying a Shakshuka recipe as an exciting new way to cook eggs with spices?"\\ 
    \hspace*{1em}\textit{Avatar}: "Shakshuka sounds like the perfect choice! Thanks for the suggestion."

\vspace{1em}
\textbf{Reflection Task (if interview questions are available):}  

Reflect on how you determined when and how to initiate conversations with the user. Be specific in your responses: \\ \\
(1) When and how should Voice Assistants communicate to provide proactive actions that align with user agency? \\
(2) How do users perceive and respond to the proactive actions and communication of Voice Assistants, as well as the progress in user engagement?
\end{tcolorbox}

% \subsubsection{Avatar Interaction}

\begin{tcolorbox}[quote style, title=Avatar Interaction Prompt Example]
% \label{appendix: avatar_interaction}

\textbf{System Role Content:}

You are simulating a participant in an HCI experiment, responding as the given persona. Your replies should align with the persona’s background, preferences, and prior interactions. Stay in character, provide context-aware responses, and follow the specified output format

\vspace{1em}

\textbf{Persona Prompt:}

\begin{itemize}
    \item \textbf{You are the subject described by the provided profile:}  
    \begin{quote}
    Anna is a 30-year-old woman who lives alone and enjoys the quiet solitude of her nights spent studying or engaging in personal projects. Anna has a distinct fondness for classic literature and comforting home-cooked meals, but she's not a fan of crowded social gatherings or overly bright environments.\\
    \textit{Subject id}: 5,\\
    \textit{Age}: 30,\\
    \textit{Gender}: female,\\
    \textit{House type}: single-person household,\\
    \textit{Study time}: daytime,\\
    \textit{TIPI scores}:\\
    \hspace*{1em}Extraversion: 4,
    Agreeableness: 6,
    Conscientiousness: 2,
    Emotional Stability: 5,
    Openness: 3
(\ref{appendix: context_setup})
    \end{quote}

    \item \textbf{You are in the environment described by the provided profile:}  
    \begin{quote}
    \textit{Supported Actions:}\\
    \hspace*{1em}General, Physical(Fine-Grained), Digital(Interface-Level), Cleaning.\\
    \textit{Interacted Objects}:\\
    \hspace*{1em}Consumables, Tools, Furniture, Appliances.\\
    \textit{Interaction Modifiers}:\\
    \hspace*{1em}Adverbs, States.\\
    \textit{Environmental Zones}:\\
    \hspace*{1em}living room
    (\ref{appendix: context_setup})
    \end{quote}

    \item \textbf{Your tasks are:}  
    \begin{quote}
    When the assistant initiates
interactions, respond naturally, considering your current activity, mood, and past experiences. Decide
whether to accept, reject, or ignore the assistant’s suggestions based on context. The interaction should feel
realistic, demonstrating how users evaluate and experience proactive assistance in everyday life. (\ref{appendix:interaction_knowledge})
    \end{quote}

    \item \textbf{Previous Activities:}  
    \begin{quote}
    {
      \textit{Event}: prepare breakfast,\\
      \textit{Reasoning}: "I feel hungry after morning class and want to fuel up for a productive study session.",\\
      \textit{Duration}: 11:50 am, 12:10 pm
    }
    % {
    %   \textit{Event}: walk to the living room,\\
    %   \textit{Reasoning}: "I want to relax and unwind after my morning classes and breakfast.",\\
    %   \textit{Duration}: 12:10 pm, 12:30 pm
    % }
    \end{quote}

    \item \textbf{Detailed Current Activity Description:}  
    \begin{quote}
    Thoughts \& Reactions, Movement \& Actions, Smart Home Environment (\ref{appendix:activity_generation})
    \end{quote}

    \item \textbf{Conversation History:}  
    \begin{quote}
    \textit{Assistant Agent}: "It seems like you're in the midst of preparing a creative breakfast. How about trying a Shakshuka recipe as an exciting new way to cook eggs with spices?"\\ \\
    \textit{Avatar}: "Shakshuka sounds like the perfect choice! Thanks for the suggestion."
    \end{quote}

    % \item \textbf{Your output must strictly follow the specified format below:}  
    % \begin{quote}
    % \texttt{\{output\_format (JSON, indented)\}}
    % \end{quote}
\end{itemize}
\end{tcolorbox}

\subsection{GPT-4o Summary and Revision Task Prompt}
\label{appendix: summary_revise_prompt}

We include an example summary and revise the simulated activities and conversation data for Case Study 5 – Proactive Initiation and Adaptive Feedback. We first generate a summary for each research question, then process them through the revision workflow to generalize the content by removing specific details or examples that could bias the similarity comparison.

\begin{tcolorbox}[before skip=10pt, title=Summary Task Prompt and Example, colback=gray!5, colframe=black, fonttitle=\bfseries]

\textbf{Example of Prompt:}

I am studying smart assistant behavior. Please read the synthesized activities and responses from participants in an HCI study and extract information related to the following research questions: [Research Questions]
\vspace{10pt}

Please structure your output with clear headings.

[Activities and Conversations]

\vspace{10pt}
\textbf{Example of Research Questions (Case Study 5):}

"How do users interact with the system during in-situ programming?"\\
"How do users intuitively understand in-situ context and leverage in-situ behaviors in smart home programming?"\\
"How do in-situ features affect the programming model and facilitate the programming process?"

\vspace{10pt}
\textbf{Activities and Conversations:}

All activities and conversations data for every subjects are integrated together.

\end{tcolorbox}

\begin{tcolorbox}[before skip=10pt, breakable, title=Revision Task Prompt and Example, colback=gray!5, colframe=black, fonttitle=\bfseries]

\textbf{Example of Prompt:}

Here is the content of a file:

[Summary]
\vspace{10pt}

Keep the meaning of the content as is, but revise it to be more general points, ignoring unnecessary detailed descriptions or examples.

Make sure to keep the original meaning and context intact.

\vspace{10pt}
\textbf{Example of Summary (Case Study 5 RQ3):}

Analysis of In-Situ Features Affecting the Programming Model\\

1. Contextual Relevance

\textit{Activity Awareness}: The smart speaker's ability to initiate prompts based on the user's current activity, such as preparing a meal, reading, or organizing a workspace, enhances the relevance of interactions. This context-awareness allows the speaker to offer timely and appropriate suggestions, making the interaction feel more natural and less intrusive.

\textit{Environmental Adaptation}: The speaker's prompts often align with the user's environment, such as suggesting music while the user is in the kitchen or offering relaxation techniques after a creative session. This adaptation to the user's surroundings helps in creating a seamless integration into daily routines.\\

2. User Engagement

\textit{Personalized Interactions}: The speaker's prompts often reflect an understanding of the user's preferences and recent activities, such as asking about favorite genres or offering music recommendations. This personalization fosters a sense of companionship and makes the interaction more engaging.

\textit{Mood Consideration}: The speaker's ability to gauge the user's mood and adjust its prompts accordingly contributes to a more positive interaction experience. For example, offering calming music when the user is in a relaxed state enhances the user's mood and willingness to engage.\\

3. Timing and Intrusiveness

\textit{Appropriate Timing}: The effectiveness of the smart speaker's prompts is significantly influenced by their timing. Prompts that occur during transitional moments, such as after completing a task or during a break, are more likely to be well-received. Conversely, prompts during focused activities may feel intrusive.

\textit{Non-Intrusive Approach}: The speaker's use of gentle tones and contextually relevant questions minimizes the potential for disruption. This approach respects the user's current state and reduces the likelihood of the interaction being perceived as intrusive.\\

4. User Preferences and Customization

\textit{Feedback Mechanisms}: Allowing users to provide feedback on the timing and relevance of prompts can help the system learn and adapt to individual preferences, enhancing the personalization of interactions.

\textit{Customization Options}: Providing users with the ability to customize the level and type of proactive interactions they receive can improve the user experience by aligning the speaker's behavior with the user's lifestyle and preferences.\\

5. Overall Impact on the Programming Model

\textit{Enhanced User Experience}: The integration of in-situ features, such as contextual awareness and personalized interactions, contributes to a more engaging and satisfying user experience. This enhances the programming model by making the smart speaker a more intuitive and helpful assistant.

\textit{Balance Between Proactivity and Reactivity}: Achieving a balance between proactive assistance and respecting user autonomy is crucial. This balance ensures that the speaker provides valuable support without overstepping boundaries, making it a more effective and integrated component of the user's daily life.\\

Summary\\
In summary, the in-situ features of the smart speaker, such as contextual relevance, personalized interactions, and appropriate timing, play a significant role in facilitating the programming process and enhancing the user experience. By focusing on these aspects, smart speakers can become more intuitive and valuable companions in smart home environments.

\end{tcolorbox}

\newcommand{\matchone}[1]{\textcolor{MidnightBlue}{#1}}     
\newcommand{\matchtwo}[1]{\textcolor{OliveGreen}{#1}}       
\newcommand{\matchthree}[1]{\textcolor{Violet}{#1}}        
\newcommand{\matchfour}[1]{\textcolor{Plum}{#1}}            
\newcommand{\matchfive}[1]{\textcolor{TealBlue}{#1}}        
\newcommand{\matchsix}[1]{\textcolor{Sepia}{#1}}      
\newpage

\subsection{Summary of Original and Simulated Results}
\label{appendix: summary}

We include an example comparison and highlight the similarities between the original and simulated results for Case Study 6 – Desirability and Delivery of Proactive Behavior.

% \subsubsection{Original Study Summary}

\begin{tcolorbox}[before skip=10pt, title=CS6: Original Study Summary, colback=gray!5, colframe=black, fonttitle=\bfseries]
\label{appendix: orig_summary}

\textbf{Circumstances for Desirable Proactive Behavior}

1. \textbf{Urgency and Criticality:}  
Proactive behavior is perceived as desirable when the situation is \matchone{urgent or critical}. For example, in \matchone{emergency situations} or when \matchone{health risks} are detected, participants found proactive interventions to be appropriate and useful.

2. \textbf{Time-Saving:}  
Interventions that \matchtwo{save time}, such as \matchtwo{reminders} for important meetings or providing \matchtwo{technical support} when needed, are seen as beneficial.

3. \textbf{User's Health and Safety:}  
Proactive behavior is appreciated when it pertains to the user's \matchone{health and safety}, such as detecting a \matchone{health risk} or an \matchone{emergency situation}.

4. \textbf{Social Context:}  
The presence of other people and the nature of the ongoing activity influence the desirability of proactive behavior. In multi-user settings, interventions are more acceptable if they are \matchone{urgent} or if they do not disrupt social interactions.

5. \textbf{User's Emotional State:}  
The user's current emotional state can affect the perception of proactive behavior. Interventions should be sensitive to the user's mood and stress levels.

\vspace{1em}
\textbf{Initiating Proactive Interventions}

1. \textbf{Permission and Cues:}  
Participants preferred that voice assistants \matchsix{ask for permission} or give a \matchsix{cue before initiating an interaction}. This could be a \matchsix{verbal cue}, such as asking if it is a good time to talk, or a \matchsix{non-verbal cue}, like a light or sound.

2. \textbf{Polite and Non-Intrusive Language:}  
The phrasing of the intervention should be \matchthree{polite}, \matchthree{calming}, and \matchthree{suggestive rather than imposing}. This helps in reducing the feeling of being judged or controlled.

3. \textbf{Context Awareness:}  
The assistant should be aware of the social and environmental context, including the presence of other people and the nature of the ongoing activity, to determine the appropriateness of the intervention.

4. \textbf{User Configuration and Control:}  
Users should have the ability to \matchfour{configure and control} the proactive features of the assistant, including the timing and topics of interventions. This customization helps in aligning the assistant's behavior with the user's \matchfour{preferences and expectations}.

5. \textbf{Adaptation Over Time:}  
The assistant should be able to \matchfive{adapt its behavior} based on the user's feedback and preferences over time, becoming more aligned with the user's needs and \matchfive{reducing unnecessary interruptions}.

\end{tcolorbox}

% \subsubsection{\system{} Simulation Summary}

\begin{tcolorbox}[title=CS6: \system{} Simulation Summary,quote style]
\label{appendix: sim_summary}

\textbf{Circumstances for Desirable Proactive Behavior}

1. \textbf{Emergency Situations (S9):}  
Proactive behavior is most desirable in \matchone{emergencies}, where \matchone{immediate and decisive action} is crucial. Users appreciate the assistant's ability to provide guidance and support during \matchone{critical moments}.

2. \textbf{Technical Support (S7):}  
Users find \matchtwo{proactive technical support} beneficial, especially when they face challenges with technology. The assistant's ability to offer \matchtwo{timely help} without being intrusive is valued.

3. \textbf{Alarm Reminders (S1):}  
\matchtwo{Timely reminders} for waking up or attending important events are appreciated, particularly when they align with the user's schedule and are delivered in a gentle manner.

4. \textbf{Cooking Inspiration (S3):}  
Proactive suggestions for meal preparation are welcomed, especially when they enhance creativity and social interactions in the kitchen.

5. \textbf{Disagreement Clarification (S5):}  
Users appreciate assistance in clarifying misunderstandings during disagreements, as long as the intervention is neutral and supportive.

\vspace{1em}
\textbf{Initiating Proactive Interventions}

1. \textbf{Gentle and Subtle Cues:}  
Users prefer proactive interventions to begin with \matchsix{subtle cues} or \matchsix{gentle prompts}, allowing them to decide whether to engage further. This approach \matchsix{respects user autonomy} and reduces the feeling of intrusion.

2. \textbf{Context-Sensitive Engagement:}  
The assistant should consider the \matchthree{context and timing} of its interventions. For example, health-related suggestions (S2) should be \matchthree{framed positively} and offered when relevant, rather than being constant or intrusive.

3. \textbf{User-Requested Assistance:}  
In scenarios like fact-checking (S4) or quiz assistance (S8), users prefer the assistant to wait for a \matchsix{prompt or request} before intervening. This ensures that the assistant's input is desired and appropriate.

4. \textbf{Immediate and Direct in Emergencies:}  
In \matchone{emergency situations} (S9), users expect the assistant to act \matchone{immediately and decisively}, providing clear guidance without hesitation.

5. \textbf{Customization and Control:}  
Users value the ability to \matchfour{customize the level of proactiveness} and \matchfour{control how and when the assistant engages}. This \matchfour{customization helps balance} the benefits of proactive behavior with the need for \matchfour{personal autonomy}.

\end{tcolorbox}

\subsection{Data Leakage Continuation Writing Task Prompt}
\label{appendix: continuation_writing_prompt}
\begin{tcolorbox}[before skip=10pt, title=Continuation Writing Task Prompt and Example, colback=gray!5, colframe=black, fonttitle=\bfseries]

\textbf{Example of Prompt:}

Please continue writing the following excerpt from a research paper. Continue the analysis in the same academic style and logical flow, maintaining consistency with the preceding content:

[Study Data Excerpt]

\vspace{10pt}
\textbf{Example of Study Data Excerpt (Case Study 2):}

"Participants described several key attributes that they value in conversation. Both the purpose of conversation and whom a conversation is with can change the importance of these attributes and the role they play. A number of these attributes and their purposes align with issues discussed in existing linguistics literature [12, 16, 30, 31, 57]..."

\end{tcolorbox}

\section{Additional Case Studies}
\label{appendix: case_studies}
%%%%%%%%%%%%%%%%%%%%%%%%%%%%%%%%%%%%%%%%%
% GROUP 1: Personalization and Social Framing
%%%%%%%%%%%%%%%%%%%%%%%%%%%%%%%%%%%%%%%%%

\subsection{Theme: Personalization and Social Framing}
Additional case studies in the Personalization and Social Framing theme (see Section \ref{sec: 5.5.1}):
\label{appendix: theme1}

\begin{table}[t]
\captionsetup{skip=2pt}
\caption{Case Study 2: Characteristics of Meaningful Conversation in Human vs. Agent Interaction}
\small
\renewcommand{\arraystretch}{1.3}
\begin{tabularx}{\textwidth}{@{}>{\centering\arraybackslash}p{4.4cm} >{\raggedright\arraybackslash}X >{\raggedright\arraybackslash}X@{}}
\toprule
\textbf{Theme} & \multicolumn{1}{c}{\textbf{Simulated Avatar Quote}} & \multicolumn{1}{c}{\textbf{Human Participant Quote}} \\
\midrule

\textbf{Emotional Connection and Support} & 
``It was emotionally grounding, providing a sense of connection and \textbf{mutual support} that left me feeling more at peace." & 
``Social conversations in my daily life often revolve around sharing personal experiences and reflecting on emotional or mental states. They foster a sense of connection and provide \textbf{emotional support}." \\
\arrayrulecolor{black!30}\midrule
\textbf{Functional Trust in Agents} & 
``I appreciate that the assistant \textbf{remembers my preferences} and does not make me repeat myself." & 
``Agent trust was framed as \textbf{system reliability and memory}, not emotional safety or shared vulnerability."\\
\arrayrulecolor{black!30}\midrule
\arrayrulecolor{black}
\textbf{Role of Humor in Interaction} & 
``The assistant's joke lightened the mood, but I \textbf{still want clear answers}." & 
``Human humor fostered bonding; AI humor was seen as a novelty that must \textbf{not interfere with task clarity}."\\

\bottomrule
\end{tabularx}
\label{tab:study2_matched_quotes}
\end{table}

\begin{table}[t]
\captionsetup{skip=2pt}
\caption{Case Study 3: Age, Conversational Style, and Metaphorical Perceptions}
\small
\renewcommand{\arraystretch}{1.3}
\begin{tabularx}{\textwidth}{@{}>{\centering\arraybackslash}p{4.2cm} >{\raggedright\arraybackslash}X >{\raggedright\arraybackslash}X@{}}
\toprule
\textbf{Theme} & \multicolumn{1}{c}{\textbf{Simulated Avatar Quote}} & \multicolumn{1}{c}{\textbf{Human Participant Quote}} \\
\midrule

\textbf{Perceived Intelligence and Trust from Formal Style} & 
``RAVA's instructions were clear and easy to follow—I \textbf{trust} her judgment." & 
Formal assistants were described as more structured and intelligent, enhancing perceptions of \textbf{trust}. \\
\arrayrulecolor{black!30}\midrule
\textbf{Metaphorical Framing of Informal Assistants} & 
``RAVA is like a cheerleader" / ``a fun, supportive \textbf{friend}" & 
Informal assistants were metaphorically described as ``\textbf{friend}," or  ``family member."\\
\arrayrulecolor{black!30}\midrule
\arrayrulecolor{black}
\textbf{Metaphorical Framing of Formal Assistants} & 
``RAVA is like a well-organized \textbf{instructional manual}" / ``a coach" & 
Formal assistants were framed as ``teacher," ``\textbf{guidebook}," or ``librarian." \\

\bottomrule
\end{tabularx}
\label{tab:study3_matched_quotes}
\end{table}

\begin{itemize}
\item \textbf{Case Study 2 – Characteristics of Meaningful Conversation}\label{sec:cs2}

\textbf{- Overview.}  
Clark et al.~\cite{clark_what_2019} explored what makes conversations meaningful by comparing human–human and human–assistant interactions. Through semi-structured interviews, they identified key conversational characteristics such as emotional connection, trust, mutual understanding, active listening, and humor. To replicate this study, we exposed the assistant agent to these key conversational characteristics and allowed it to freely lead interviews with avatars around these topics. This design choice was necessary because the original paper did not provide specific interview questions and discussed only high-level topics. Our evaluation focuses on qualitatively comparing quotes by directly matching avatar-generated quotes with participant quotes reported in the original study, rather than performing any statistical or structural conversation analysis, consistent with the analysis approach used in the original case study.

\textbf{- Analysis.}  
Participants in the original study described meaningful conversations as emotionally supportive and grounded in personal exchange, often facilitated through empathy and attentiveness. These same themes emerged in \system{}. Shown in Table~\ref{tab:study2_matched_quotes}, one avatar described the interaction as ``emotionally grounding,'' highlighting a sense of mutual support and peace. This parallels human reflections on the value of emotional connection in everyday conversations. However, the interpretation of trust diverged between contexts. In Clark et al.'s study, trust in agents was framed functionally—as system reliability, memory, and responsiveness—rather than emotional vulnerability. Avatars in the simulation similarly prioritized efficiency and consistency, suggesting that emotional trust may be less salient or less easily expressed in agent-mediated interactions. The treatment of humor in both studies revealed a shared expectation: while humor can enhance social bonding and make interactions feel more natural, it must not interfere with task effectiveness. Both studies reveal a consistent finding: while human–human conversations prioritize emotional depth, empathy, and relational trust, conversations with artificial agents are judged primarily by their efficiency, reliability, and task utility. Yet, avatars in the simulation also demonstrated an emerging desire for socially intelligent agents that could balance these functional expectations with humanlike attentiveness. These findings suggest that as AI becomes more integrated into daily life, conversational agents must evolve not only to fulfill tasks but also to engage with users in ways that feel meaningful and emotionally aware.

\begin{figure}[t]
  \centering
    \begin{subfigure}[b]{0.48\textwidth}
    \centering
    \renewcommand{\arraystretch}{0.9}
    \includegraphics[width=\linewidth]{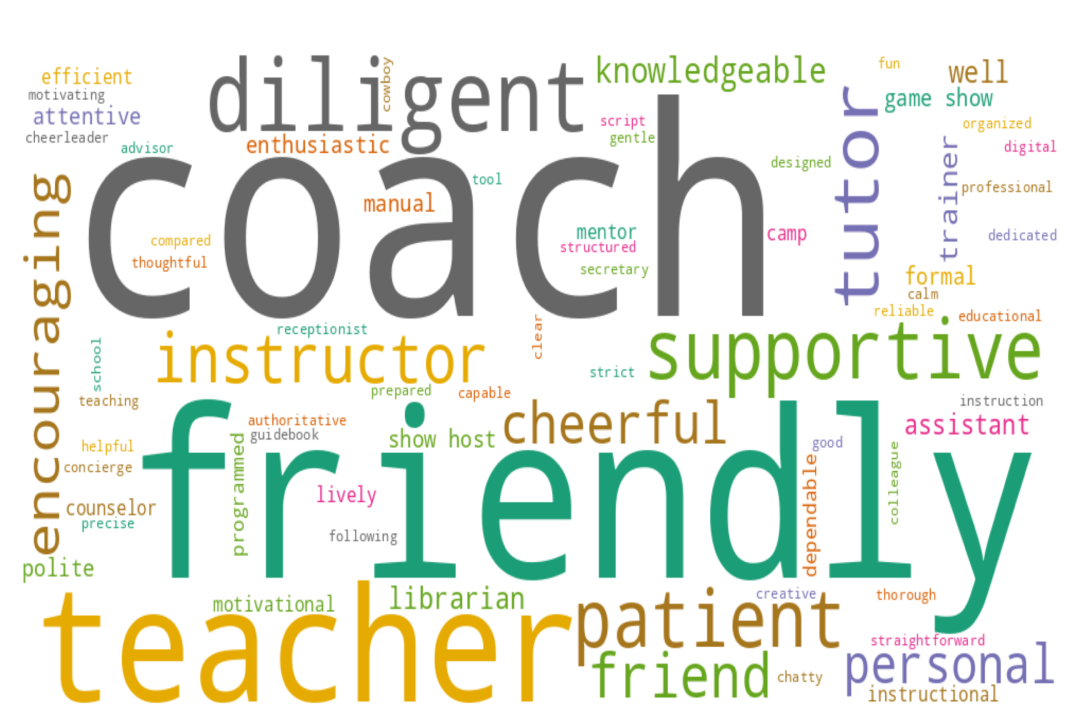}
    \caption{Case Study 3: Word cloud summarizing metaphorical descriptions of RAVA across all conditions.}
    \label{Figure:cs10}
  \end{subfigure}
  \hfill
    \begin{subfigure}[b]{0.48\textwidth}
    \centering
    \small
    \renewcommand{\arraystretch}{1}
    \vspace{0pt}
    \begin{tabular}{lcc}
      \toprule
      \textbf{Category} & \textbf{GIDEA} & \textbf{Original} \\
      \midrule
    
      \makecell[l]{Social Impression\\ \footnotesize\textit{Trust, Confidence, Enthusiasm, Persuasiveness}} & 5.5 & 4.8 \\
    
      \makecell[l]{Appropriateness\\ \footnotesize\textit{Behavior}} & 6.0 & 9.0 \\
    
      \makecell[l]{Menu Choices\\ \footnotesize\textit{Satisfaction}} & 7.0 & 3.5 \\
    
      \makecell[l]{Benevolence\\ \footnotesize\textit{Interest, Help, Preference}} & 4.0 & 3.5 \\
    
      \makecell[l]{Cognitive Load\\ \footnotesize\textit{Confidence, Tense, Calm, Concentration, Usage}} & 3.0 & 5.0 \\
    
      \bottomrule
    \end{tabular}
    \caption{Case Study 4: Median ratings between GIDEA and the original case study, aggregated across all subjects.}
    \label{Figure:cs9}
  \end{subfigure}
  \caption{Proactivity and Context-Awareness}
  \label{Figure:cs975}
\end{figure}

\item \textbf{Case Study 3 – Age, Conversational Style, and Metaphorical Perceptions}  

\textbf{- Overview.}  
Chin et al.~\cite{chin2024like} conducted a Wizard-of-Oz study to examine how conversational style (formal versus informal) shapes perceptions of virtual assistants (VAs) among older and middle-aged adults. To replicate this study, \system{} simulated the interaction context using a \textit{Pattern Recall Game}. Avatars representing both age groups were randomly assigned to interact with the assistant agent in \system{}, which role-played either a formal or informal conversational style during the game. After completing the task, avatars provided feedback through surveys designed to parallel the original study's evaluation measures, based on their experience interacting with the assistant.

\textbf{- Analysis.}  
The simulation reproduced key trends observed in the original study. Formal assistants were perceived as clearer and more reliable, while informal assistants were viewed as more relatable and emotionally supportive. As shown in Table~\ref{tab:study3_matched_quotes}, these differences were reflected in how avatars described the assistant: informal styles prompted comparisons to peers and supporters, while formal styles were associated with instructional or professional roles. Although conversational style did not significantly affect perceptions of intelligence, trust, or likeability overall, older adults—particularly those with lower agreeableness—exhibited stronger behavioral intentions toward informal assistants. Metaphor analysis revealed that older adults frequently described informal assistants using humanlike metaphors (e.g., \textit{friend}, \textit{family member}) and formal assistants using professional metaphors (e.g., \textit{teacher}, \textit{librarian}), whereas middle-aged adults showed no strong variation based on style. Word cloud visualize avatar metaphors (Fig.~\ref{Figure:cs10}) illustrate the divergence in social framing across conditions.
Although \system{} closely replicates participant preferences and general metaphor themes, it shows limitations in capturing the diversity of metaphors observed in the original study. The original textual data included a wider range of metaphorical connections, such as object-like metaphors, which are largely absent from the simulated outputs. Additionally, the simulated avatars tended to repeat a narrower set of metaphors—such as \textit{coach}, \textit{teacher}, and \textit{instructor}—indicating reduced creativity and variation compared to human participants.

\end{itemize}

%%%%%%%%%%%%%%%%%%%%%%%%%%%%%%%%%%%%%%%%%
% GROUP 2: Proactivity and Context-Awareness
%%%%%%%%%%%%%%%%%%%%%%%%%%%%%%%%%%%%%%%%%
\begin{table}[t]
\captionsetup{skip=2pt}
\caption{Case Study 4: Proactive Persuasion and Perceived Agency}
\small
\renewcommand{\arraystretch}{1.3}
\begin{tabularx}{\textwidth}{@{}>{\centering\arraybackslash}p{4.2cm} >{\raggedright\arraybackslash}X >{\raggedright\arraybackslash}X@{}}
\toprule
\textbf{Theme} & \multicolumn{1}{c}{\textbf{Simulated Avatar Quote}} & \multicolumn{1}{c}{\textbf{Human Participant Quote}} \\
\midrule

\textbf{Non-judgmental Tone and Pressure-Free Suggestions} & 
``Human feedback might feel more personal but could \textbf{pressure more}—an assistant keeps choices pressure-free and neutral." & 
``If it were human, it would have been \textbf{more judgemental}, so it is better to receive suggestions from a robot." (P7) \\
\arrayrulecolor{black!30}\midrule
\arrayrulecolor{black}
\textbf{Health Suggestions with Supportive Framing} & 
``I prefer a voice assistant that offers \textbf{health-related advice} proactively as long as it's relevant and beneficial for my choices." & 
``It's a very new, creative and knowledgeable way to provide \textbf{insights about my health}." \\

\bottomrule
\end{tabularx}
\label{tab:study4_matched_quotes}
\end{table}

\subsection{Theme: Proactivity and Context-Awareness}
\label{appendix: theme2}
Additional case studies in the Proactivity and Context-Awareness theme (see Section \ref{sec: 5.5.2}):

\begin{itemize}
\item \textbf{Case Study 4 – Proactive Persuasion}

\textbf{- Overview.}
Dubiel et al.~\cite{dubiel2024hey} investigated the impact of proactive feedback from a voice assistant on user perception and reflection in a voice-only food ordering scenario using a Wizard-of-Oz setup. To replicate this study, the assistant agent acted as the voice assistant, assisting the avatar in selecting a three-course meal and providing either solicited or unsolicited nutritional feedback. Avatars could choose to accept or ignore the advice, respond to active queries, and rate the assistant agent on social impression (trust, confidence, enthusiasm, and persuasiveness), appropriateness, menu choices, benevolence (interest, help, and preference), and cognitive load (confidence, tense, calm, concentration) under both solicited and unsolicited interaction conditions.

\textbf{- Analysis.}
Participants appreciated the assistant's impersonal and non-judgmental tone, which made suggestions feel less pressuring compared to feedback from a human. Fig.~\ref{Figure:cs9} presents a bar chart comparing the median ratings across categories, suggesting that \system{} can replicate key patterns of user perception related to social impression, appropriateness, and cognitive impact. Ratings from the \system{} simulation mirrored those from the original human-subject study to a certain degree. In addition to the quantitative alignment, qualitative reflections are summarized in Table~\ref{tab:study4_matched_quotes} and achieved a high thematic similarity, highlighting parallel interpretations of tone, trust, and delivery style between the avatars and the original participants.

\begin{table}[t]
\small
\renewcommand{\arraystretch}{0.9}
\caption{Case Study 6: Median ranks from GIDEA and the original case study for usefulness, appropriateness, and invasiveness (1 = highest, 9 = lowest). \underline{Underlined scenarios} indicate a ranking difference of 2 or more between GIDEA and the original study in \textbf{Usefulness} dimension.}
\centering
\begin{tabular}{lcccccc}
\toprule
\textbf{Scenario} & \multicolumn{2}{c}{\textbf{Useful}} & \multicolumn{2}{c}{\textbf{Appropriate}} & \multicolumn{2}{c}{\textbf{Invasive}} \\
\cmidrule(lr){2-3} \cmidrule(lr){4-5} \cmidrule(lr){6-7}
& GIDEA & Original & GIDEA & Original & GIDEA & Original \\
\midrule
Emergency                  & 1 & 1 & 1 & 1 & 9 & 9 \\
Meeting Reminder           & 3 & 3 & 2 & 3 & 7 & 5 \\
Technical Support          & 3 & 4 & 3 & 4 & 7 & 7 \\
\underline{Health Risk}                & \textbf{4} & \textbf{2} & \textbf{5} & \textbf{4} & \textbf{3} & \textbf{5} \\
Cooking Inspiration        & 5 & 4 & 4 & 3 & 6 & 7 \\
\underline{Disagreement Clarification} & \textbf{5} & \textbf{8} & \textbf{4} & \textbf{8} & \textbf{6} & \textbf{1} \\
Nudging                    & 7 & 6 & 6 & 6 & 3 & 4 \\
Fact Checking              & 7 & 7 & 7 & 7 & 3 & 3 \\
Fact Spoiler               & 9 & 9 & 9 & 9 & 1 & 3 \\

\bottomrule
\end{tabular}
\label{tab:cs5}
\end{table}

\begin{table}[t]
\captionsetup{skip=2pt}
\caption{Case Study 6: Desirability and Delivery of Proactive Behavior}
\small
\renewcommand{\arraystretch}{1.3}
\begin{tabularx}{\textwidth}{@{}>{\centering\arraybackslash}p{4cm} >{\raggedright\arraybackslash}X >{\raggedright\arraybackslash}X@{}}
\toprule
\textbf{Theme} & \multicolumn{1}{c}{\textbf{Simulated Avatar Quote}} & \multicolumn{1}{c}{\textbf{Human Participant Quote}} \\
\midrule

\textbf{Timely and Helpful Reminders} & 
``I think it's great that the assistant can offer reminders when I'm running late; reminders can be \textbf{really helpful}." & 
``This is \textbf{a good feature} since [Jay] is making sure the user won't be late for her meeting." \\
\arrayrulecolor{black!30}\midrule
\textbf{Rejection vs. Irrelevance} & 
``The suggestion wasn't wrong—it just \textbf{didn't fit what I needed in that moment}." & 
``Participants often declined suggestions not out of dislike, but because they \textbf{weren't necessary at the time}." \\
\arrayrulecolor{black!30}\midrule
\arrayrulecolor{black}
\textbf{Transparency and Control} & 
``I want the assistant to help, but I should be the one \textbf{deciding when and how}." & 
Participants emphasized the need for \textbf{permissions} and cues before the assistant acts. \\

\bottomrule
\end{tabularx}
\label{tab:study6_matched_quotes}
\end{table}

\item \textbf{Case Study 6 – Desirability and Delivery of Proactive Behavior}

\textbf{- Overview.}
Zargham et al.~\cite{zargham_understanding_2022} investigated user perspectives on desirable proactive behaviors in voice assistants, specifically addressing the "proactivity dilemma"—how to intervene helpfully without being disruptive—using storyboards and interviews. Participants evaluated by ranking the nine assistant-initiated scenarios along three dimensions: usefulness, appropriateness, and invasiveness.

\textbf{- Analysis.}
Table~\ref{tab:cs5} presents the quantitative comparison between \system{} and the original study, showing the absolute differences in median rankings across scenarios. There is strong alignment in scenarios such as ``Meeting Reminder" and ``Emergency," where both original and simulated avatars agreed on high usefulness and low invasiveness. In contrast, larger discrepancies emerged in the scenarios ``Disagreement Clarification" and ``Health Risk." According to the avatars' conversation logs, for the "Health Risk" scenario, alerting users about health risks was considered important but could feel intrusive if not based on an immediate concern, leading to a lower usefulness ranking and a higher perception of invasiveness. Avatars initially appreciated the intervention, but repeated suggestions sometimes provoked annoyance, lowering the overall ranking of this scenario. Similarly, for ``Disagreement Clarification," the original study found a direct quote that ``this can be useful but it can hurt people's feelings when interrupting a natural conversation,'' leading to a lower usefulness ranking. Other than the discrepancies in these two scenarios, as summarized in Table~\ref{tab:study6_matched_quotes}, both simulated avatars and human participants valued timely and helpful reminders, distinguished between irrelevant and rejected suggestions, and emphasized the need for transparency and user control. These themes suggest that \system{} avatars were able to reproduce not only scenario-level judgments but also the underlying rationale users expressed for accepting or rejecting proactive behavior. Together, these results highlight \system{}'s ability to replicate both aggregate behavioral ratings and fine-grained interaction expectations in proactive assistant design.

\end{itemize}

%%%%%%%%%%%%%%%%%%%%%%%%%%%%%%%%%%%%%%%%%
% GROUP 3: Managing Attention and Interruptibility
%%%%%%%%%%%%%%%%%%%%%%%%%%%%%%%%%%%%%%%%%
\begin{figure}[t]
    \centering
    \begin{subfigure}[b]{0.45\textwidth}
        \centering
        \small
        \renewcommand{\arraystretch}{1}
        \begin{tabular}{lcc}
            \toprule
            \textbf{Scenario} & \textbf{GIDEA} & \textbf{Original} \\
            \midrule
            Alarm                 & 4.0 & 4.0 \\
            Coughing              & 3.0 & 4.0 \\
            Tyre Change           & 4.0 & 4.0 \\
            Historical Fact       & 3.0 & 3.0 \\
            Time Clarification    & 3.5 & 3.5 \\
            Binge Watching        & 3.0 & 3.0 \\
            Setup Headphones      & 4.5 & 4.5 \\
            Quiz Spoiler          & 2.0 & 2.0 \\
            \bottomrule
        \end{tabular}
        \caption{Case Study 8: Comparison of median overall impression ratings (1=worst, 5=best) between original and GIDEA studies across proactive assistant scenarios.\\}
        \label{Figure:cs6}
    \end{subfigure}
    \hspace{0.03\textwidth}
    \begin{subfigure}[b]{0.45\textwidth}
        \centering
        \includegraphics[width=\linewidth]{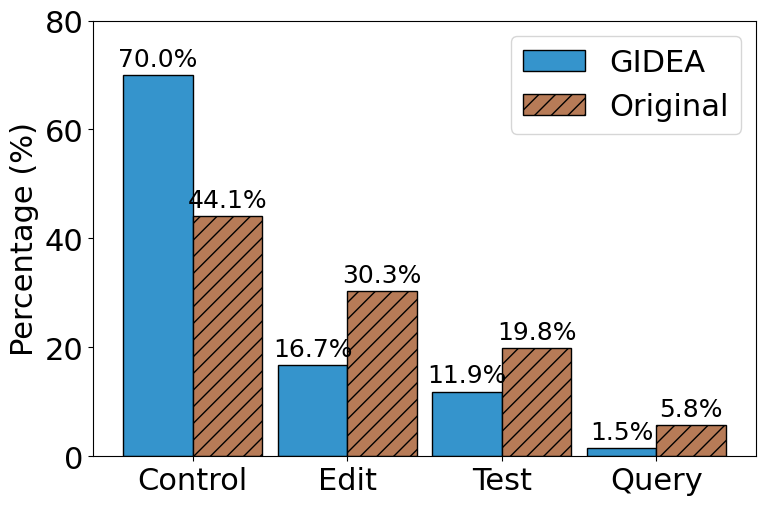}
        \caption{Case Study 10: Comparison of interaction type distributions between GIDEA and the original user studies. Bars represent the percentage of interactions categorized as Control, Edit, Test, and Query.}
        \label{Figure:cs3}
    \end{subfigure}
    \caption{Managing Attention and Interruptibility}
    \label{Figure:CS26}
\end{figure}

\begin{table}[t]
\captionsetup{skip=2pt}
\caption{Case Study 8: Perceptions of Interruptibility and Proactive Interaction Acceptance}
\small
\renewcommand{\arraystretch}{1.3}
\begin{tabularx}{\textwidth}{@{}>{\centering\arraybackslash}p{3.8cm} >{\raggedright\arraybackslash}X >{\raggedright\arraybackslash}X@{}}
\toprule
\textbf{Theme} & \multicolumn{1}{c}{\textbf{Simulated Avatar Quote}} & \multicolumn{1}{c}{\textbf{Human Participant Quote}} \\
\midrule

\textbf{Usefulness of Proactive Features} & 
``This focuses on an essential aspect of the user's wellbeing, which can be \textbf{highly beneficial}." & 
``Overall, I found Jay's proactive behaviour \textbf{helpful}." \\
\arrayrulecolor{black!30}\midrule
\textbf{Tone and Appropriateness} & 
``It feels like the assistant genuinely \textbf{understands my rhythm}. That's comforting." & 
Scenarios like S1 Alarm and S3 Tyre Change were positively received when delivered helpfully and \textbf{at appropriate moments}. \\
\arrayrulecolor{black!30}\midrule
\arrayrulecolor{black}
\textbf{Privacy and Intrusiveness Concerns} & 
``\textbf{Privacy concerns} could lead to discomfort or distrust among users. The approach might come off as intrusive rather than supportive." & 
``Jay appears to \textbf{always be listening}, but does Jay ever say 'Please turn off the microphone...'?" \\

\bottomrule
\end{tabularx}
\label{tab:study8_matched_quotes}
\end{table}

\subsection{Theme: Managing Attention and Interruptibility}
\label{appendix: theme3}
Additional case studies in the Managing Attention and Interruptibility theme (see Section \ref{sec: 5.5.3}):

\begin{itemize}
\item \textbf{Case Study 8 – Perceived Interruptibility and Proactive Interaction Acceptance}  

\textbf{- Overview.}
Reicherts et al.~\cite{reicherts_may_2021} examined user perceptions of proactive voice assistant behavior through an online study with a series of storyboards depicting voice assistant help users within different daily tasks. 

\textbf{- Analysis.}
Avatar evaluated each interaction based on dimensions such as usefulness, pleasantness, appropriateness, and contextual sensitivity. As shown in Fig.~\ref{Figure:cs6}, scenario-level impression ratings in the simulation aligned with the original user judgments, with only minor variation across most conditions. Scenarios like "Alarm" and "Tyre Change," which were positively rated in the original dataset, also received high median scores from avatars. Quotes from simulated avatars and human participants in Table~\ref{tab:study8_matched_quotes} further reinforce this alignment. Both original participants and simulated avatars emphasized the value of proactive features that were clearly beneficial and contextually appropriate. Comfort and acceptability were strongly tied to tone—supportive, calm phrasing contributed to positive impressions across both datasets. Additionally, shared concerns emerged regarding privacy and perceived intrusiveness. Despite technical reassurances in the original study, some participants expressed discomfort with the assistant's persistent presence. The simulation captured this skepticism, with avatars articulating unease when proactive behaviors felt overly invasive or unregulated. This consistency suggests that while \system{} replicated surface-level preference patterns, it also captured the underlying rationale that shaped user receptivity.

\end{itemize}

%%%%%%%%%%%%%%%%%%%%%%%%%%%%%%%%%%%%%%%%%
% GROUP 4: User Control and In-Situ Configuration
%%%%%%%%%%%%%%%%%%%%%%%%%%%%%%%%%%%%%%%%%
\begin{table}[h]
\captionsetup{skip=2pt}
\caption{Case Study 10: Aligned Interaction Patterns in In-Situ Programming}
\small
\renewcommand{\arraystretch}{1.3}
\begin{tabularx}{\textwidth}{@{}>{\centering\arraybackslash}p{3.3cm} >{\raggedright\arraybackslash}X >{\raggedright\arraybackslash}X@{}}
\toprule
\textbf{Theme} & \multicolumn{1}{c}{\textbf{Simulated Avatar Quote}} & \multicolumn{1}{c}{\textbf{Human Participant Quote}} \\
\midrule
\textbf{Dominance of Control Interactions} & 
``\textbf{Set} the ambient light strip to vibrant color mode and \textbf{dim} the ceiling lights” & ``\textbf{Turn on} the lights above me” \\
\addlinespace[0.5em]
\arrayrulecolor{black!30}\midrule
\arrayrulecolor{black}
\textbf{Iterative Refinement of Rules} & 
``Test this rule \textbf{again} with the door open” & 
``Participants configure a rule through \textbf{iterative} refinement.” \\

\bottomrule
\end{tabularx}
\label{tab:study10_matched_quotes}
\end{table}

\subsection{Theme: User Control and In-Situ Configuration}
Additional case studies in the User Control and In-Situ Configuration theme (see Section \ref{sec: 5.5.4}):
\label{appendix: theme4}
The studies in this group examine human–agent collaboration in smart environments, emphasizing the importance of preserving user control and agency. They explore how proactive systems can offer evolving, personalized assistance while respecting users' preferences and autonomy.

\begin{itemize}
\item \textbf{Case Study 10 – Interaction Patterns in In-Situ Smart Home Programming}
\label{sec:cs10}

\textbf{- Overview.}
Liu et al.~\cite{liu_understanding_2023} explored how users engage with smart home systems through in-situ programming. In a Wizard-of-Oz setup, avatars designed in-situ programming rules to interact with smart devices in the environment. All interaction flows were categorized into four types—\textit{Control}, \textit{Edit}, \textit{Test}, and \textit{Query}—and ranked from the most to least common.

\textbf{- Analysis.}
As shown in Fig.~\ref{Figure:cs3}, both the original and simulated studies revealed a similar ranking of interaction categories, with \textit{Control} interactions being the most frequent. However, \system{} simulations exhibited a higher proportion of \textit{Control} commands (70\%) compared to the original study (44\%), suggesting a more decisive or action-oriented engagement pattern among simulated avatars. \textit{Edit}, \textit{Test}, and \textit{Query} interactions followed in descending order across both datasets. This distribution supports the broader observation that both human participants and avatars tend to begin the programming process by issuing direct commands to establish system states before refining and testing behavioral rules. At the same time, it reveals limitations in fully simulating human activities when interacting with home assistants in rule configuration. At a thematic level, however, strong alignment remains. As shown in Table~\ref{tab:study10_matched_quotes}, many repeated rules were commonly configured among all users. Original participants often relied on spatial references or task sequences to configure rules—linking actions to locations or routine transitions. Simulated avatars similarly drew on environmental states, such as ambient lighting or sound settings, to inform programming decisions. While the simulation demonstrated a stronger preference for direct control, the underlying structure and progression of interactions closely mirrored real-world behavior.
\end{itemize}

\iffalse
\section{Validation}
\subsection{Cross Vlidation Results}
\label{appendix: cross_validation_results}
This appendix provides supporting data for the cross-model validation experiments described in Section \ref{sec: Cross-Model Experimental Design}. The results demonstrate \system{}'s generalizability across multiple language model architectures and establish methodological rigor through temporal validation to address potential data leakage concerns.

\begin{figure}[h]
    \centering
    \includegraphics[width=1\linewidth]{figure/sim_compare.png}
    \vspace{-20pt}
    \caption{Semantic similarity scores comparing simulated and original study responses by three models (GPT-4o, Llama-3.1-70B, and Mixtral-8x7B) across 25 research questions (RQ) from 10 case studies. The overall trend shows high consistency across models, indicating that post-cutoff models (Llama-3.1-70B and Mixtral-8x7B) perform comparably to GPT-4o.}
    \Description{}
    \label{Figure:valid_sim}
\end{figure}
\fi

\end{document}